\documentclass[10pt]{iopart}

\usepackage[utf8]{inputenc}
\usepackage[english]{babel}
\usepackage{hyperref}
\usepackage{amsfonts}
\usepackage{amssymb}
\usepackage{graphicx}
\usepackage[left=2cm,right=2cm,top=2cm,bottom=2cm]{geometry}
\usepackage{subfig}
\usepackage[dvipsnames]{xcolor}

\newcommand{\dbeta}{\Delta\beta}
\newcommand{\sinc}[1]{\mathrm{sinc}\left({#1}\right)}

\newcommand{\Ref}[1]{(\ref{#1})}

\begin{document}
\title[Fabrication limits of nonlinear waveguides]{Fabrication limits of waveguides in nonlinear crystals and their impact on quantum optics applications}
\author{Matteo Santandrea$^1$, Michael Stefszky$^1$, Vahid Ansari$^1$ and Christine Silberhorn$^1$}
\address{$^1$ University of Paderborn, Warburgerstr. 100, 33098 Paderborn, Germany}
\ead{matteo.santandrea@upb.de}
\begin{abstract}
Waveguides in nonlinear materials are a key component for photon pair sources and offer promising solutions to interface quantum memories through frequency conversion. 
To bring these technologies closer to every-day life, it is still necessary to guarantee a reliable and efficient fabrication of these devices. Therefore, a thorough understanding of the technological limitations of nonlinear waveguiding devices is paramount. 
In this paper, we study the link between fabrication errors of waveguides in nonlinear crystals and the final performance of such devices. 
In particular, we first derive a mathematical expression to qualitatively assess the technological limitations of any nonlinear waveguide.
We apply this tool to study the impact of fabrication imperfections on the phasematching properties of different quantum processes realized in titanium-diffused lithium niobate waveguides.
Finally, we analyse the effect of waveguide imperfections on quantum state generation and manipulation for few selected cases.
We find that the main source of phasematching degradation is the correlated variation of the waveguide's dispersion properties and suggest different possible strategies to reduce the impact of fabrication imperfections.
\end{abstract}
\submitto{\NJP}


\section{Introduction}
\label{sec:introduction}
Nonlinear optical processes enable complex manipulation of light and have been exploited extensively both in the classical and quantum regime for a wide variety of purposes, e.g. classical single- and multiple-channel frequency conversion \cite{Kumar1990, Chou1999}, optical parametric amplification \cite{Cerullo2003}, generation of squeezed states and entangled photons \cite{Pysher2008, Thyagarajan2009,Stefszky2017}, frequency conversion for single-photon detection \cite{Albota2004, Roussev2004, Vandevender2004} and to interface single photons with quantum memories \cite{Pelc2012a, Rutz2017, Maring2017}. 
Realizing nonlinear processes in integrated waveguides is fundamental in bringing quantum protocols and devices closer to every-day life \cite{Orieux2016}. 
Integrated nonlinear waveguides offer a few advantages over bulk nonlinear crystals, since they achieve a stronger nonlinear interaction by increasing the field confinement over longer lengths and can be interfaced more easily with fibre networks \cite{Montaut2017}.
Moreover, they can be integrated along with other linear and nonlinear elements to generate and manipulate different quantum states of light \cite{Krapick2013, Kruse2015, Sansoni2017,Lenzini2018arxiv}.  
However, the nonlinear properties of integrated waveguides critically depend upon their quality and any fabrication imperfection can degrade the final performance.

In the classical regime, studies have already been performed to understand the relationship between fabrication imperfections and phasematching properties.
In particular, Lim et al. \cite{Lim1990} introduced the concept of \textit{noncritically phasematched} waveguides, i.e. waveguides that are specifically designed to minimise the impact of fabrication imperfections. They also derived fabrication conditions for noncritically phasematched thin-film and slab waveguides. 
Experimentally, noncritical phasematching conditions for second harmonic generation have been investigated in annealed proton-exchanged lithium niobate (APE-LN) waveguides \cite{Bortz1994}. 
In the field of quantum optics, different studies have addressed the influence of fabrication imperfections on the generation of photon pairs through parametric down conversion (PDC) in waveguides \cite{Pelc2010,Pelc2011a,Phillips2013} and photonics crystal fibres \cite{FrancisJones2016}.
The vast majority of these analyses has investigated the connection between fabrication imperfections and maximum conversion efficiency of the system.
For quantum applications, however, other properties become more critical depending on the intended task of the device, e.g. the phasematching bandwidth or its shape. Therefore, it is important to analyse the influence of fabrication imperfections in quantum devices systems bearing in mind their specific application.

In this paper, we study the impact of fabrication imperfections on the performance of waveguides in nonlinear crystals. 
In particular, we analyse a variety of quantum processes realized in titanium in-diffused lithium niobate (Ti:LN) waveguides and show how their nonlinear performance is degraded by the presence of errors on the waveguide width.

In section \ref{subsec:mathdescription} we derive a qualitative expression that relates the length of a device to the maximum fabrication error tolerable. In section \ref{subsec:numericalanalysis} we apply this relation to estimate the effect of fabrication errors for a variety of different quantum optics processes of interest realized in titanium in-diffused lithium niobate (Ti:LN) waveguides. In section \ref{subsec:statanalysis} the effect of different types of fabrication imperfections is investigated by means of stochastic simulations and in section \ref{subsec:applications} we discuss how fabrication imperfections affect quantum state generation and manipulation. We focus our attention to the cases of squeezing generation, high-dimensional frequency bin encoding and efficient bandwidth compression of single photons. 
Finally, in section \ref{sec:discussion} we discuss the results of the simulations and describe possible ways to overcome the fabrication limits that degrade the phasematching.

\section{Qualitative model for fabrication tolerances}
\label{sec:qualitativeModel}
\subsection{Mathematical model}
\label{subsec:mathdescription}
We begin deriving a simple model describing the effect of fabrication imperfections on the efficiency of a nonlinear process. 
In particular, we assume that the fabrication imperfection is constant along the sample length. 
Consider a general three-wave mixing (TWM) process in a waveguide
\begin{eqnarray*}
\omega_3 &= \omega_2 + \omega_1\\
\Delta\beta &=\beta_3 - \beta_2 - \beta_1,
\end{eqnarray*}
where $\omega_i$ and $\beta_i$ are the frequencies and the momenta of the fields involved, respectively, and $i=1,2,3$ denotes the three interacting fields.
If the momentum mismatch $\Delta\beta$ is constant along the waveguide, an exact solution for the phasematching spectrum $\phi (\Delta\beta)$ of the process is given by \cite{Boyd2008}
\begin{equation}
\phi(\Delta\beta) \propto \frac{1}{L}\int_0^L\rme^{\rmi\Delta\beta z}\rmd z \Rightarrow \phi(\Delta\beta)\propto\mathrm{sinc}\left(\frac{\Delta\beta L}{2}\right)\mathrm{e}^{\rmi\frac{\Delta\beta L}{2}},
\label{eq:ideal_pm}
\end{equation}
where $L$ is the crystal length. In the case of quasi-phasematching, $\Delta\beta$ has to include the effect of the grating vector $\beta_{QPM} = 2\pi/\Lambda$.

The propagation constants $\beta_i = 2\pi n_i/\lambda_i$ depend on the refractive index $n_i$ seen by the light field as it propagates in the crystal.
Fields propagating in a waveguide see an \textit{effective} refractive index dependent on the local refractive index distribution $n(x,y,z)$ \cite{Suhara2003}.
If $n(x,y,z)$ does not vary along the propagation axis $z$ of the waveguide, then $n^{eff}$ is constant and the waveguide is said to be \textit{homogeneous}. This assumption simplifies the treatment of phasematching in waveguide structures and in the rest of this section we will consider this scenario. The analysis of spatially-varying $n^{eff}(z)$ will be analysed in detail in section \ref{subsec:statanalysis}.

\begin{figure}[tbp]
	\centering
    \includegraphics[width=.5\textwidth]{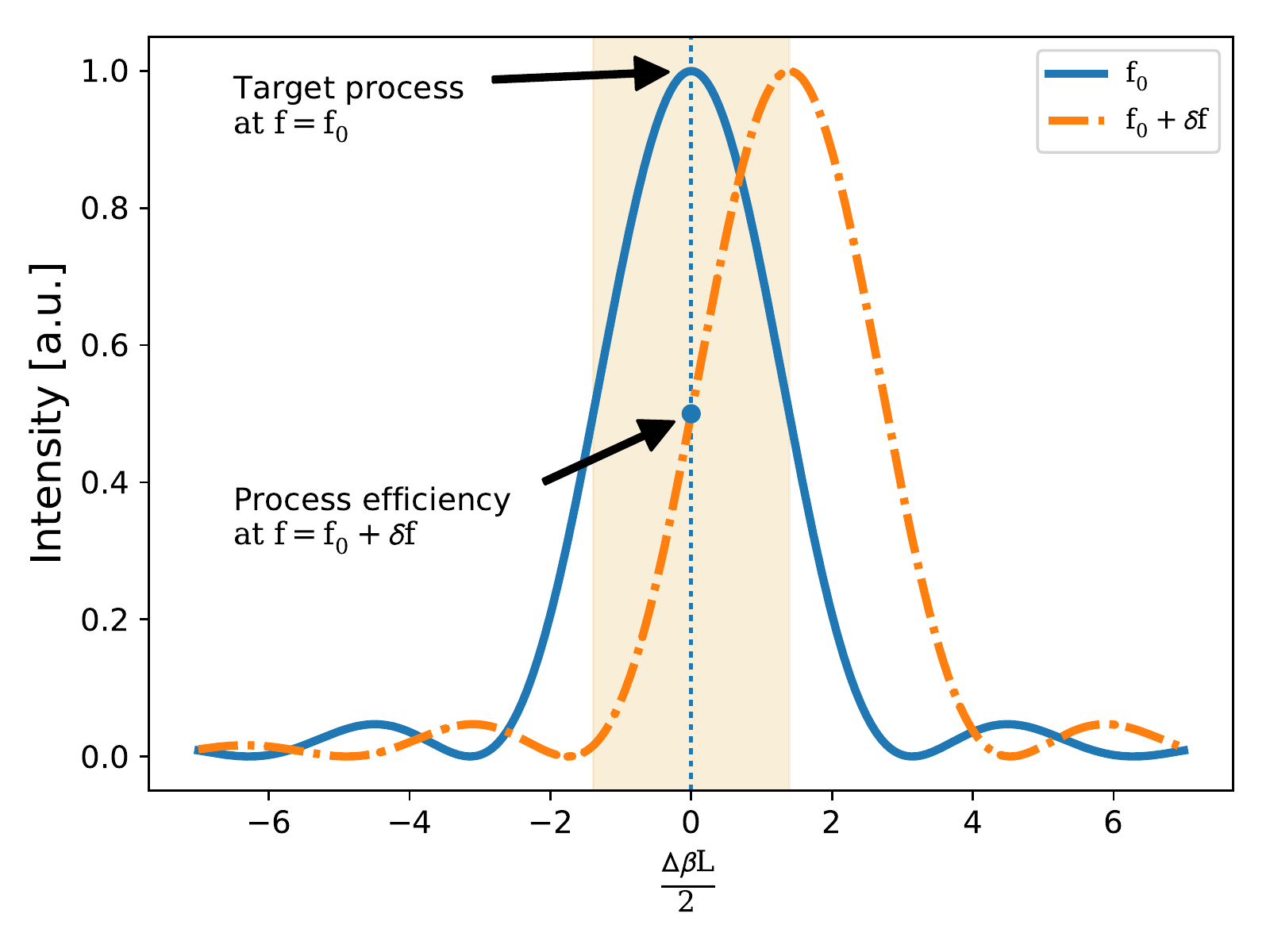}
    \caption{Illustration of the effect on the phasematching of a uniform variation $\delta f$ of the waveguide fabrication parameter $f$. The solid blue line shows the intensity of a desired nonlinear process for a waveguide with fabrication parameter $f_0$, plotted against the dimensionless parameter $\frac{\Delta\beta L}{2}$. A variation $\delta f$ of the fabrication parameter will shift the phasematching curve (dashdotted orange line). For shifts greater than the HWHM of the phasematching, the efficiency of the target process, represented by the dashed blue line, drops below 50\%. This condition is used to establish a simple criterion to indicate when the fabrication error $\delta f$ moves the process outside the chosen tolerance limits.}
	\label{img:bw_description}
\end{figure}

Following the approach presented in \cite{Lim1990}, we consider a homogeneous waveguide designed for a specific TWM process. 
For simplicity, we analyse the influence of a single fabrication parameter having a nominal value $f_0$. 
Such a parameter can represent, for example, the waveguide width, depth, exchange temperature, etc. 
Due to fabrication imperfections, the fabrication parameter $f_{prod}$ during the production can be off from the designed one by $\delta f = f_{prod}-f_0$.  The fabrication error $\delta f$ will modify the $n^{eff}$ of the waveguide, which in turn has an impact on $\Delta\beta$. For this reason, a fabrication error $\delta f$ will shift the position of the phasematching curve $\phi(\Delta\beta)$ and reduce the overall efficiency of the process, as shown in Figure \ref{img:bw_description}. 

To specify the fabrication tolerances for the waveguide production, we allow variations in $\Delta\beta$ such that the efficiency of the target process remains greater than 50\% of the ideal value:
\begin{equation}
\sinc{\frac{\Delta\beta L}{2}}^2\geq 0.5\Rightarrow \left|\frac{\Delta\beta L}{2}\right|\leq \Gamma,
\label{eq:bandwidth}
\end{equation}
where $\Gamma\approx 1.39$ is the half-width at half maximum of $\mathrm{sinc}(x)^2$.
Expanding $\Delta\beta$ in a Taylor series 
\begin{equation*}
\Delta\beta = \Delta\beta(f_0) + \left.\partial_f\Delta\beta\right|_{f_0}\cdot\delta f + o(\delta f^2),
\end{equation*}
and noticing that, for the target process, $\Delta\beta(f_0)=0$, we can approximate (\ref{eq:bandwidth}) to the first order as 
\begin{equation}
\left|\left.\partial_f\Delta\beta\right|_{f_0}\cdot\delta f\right|\frac{L}{2}\leq \Gamma.
\label{eq:sensitivity_waveguide}
\end{equation}

The parameter $\partial_f\Delta\beta$ can be referred to as the \textit{process sensitivity} to parameter $f$ because it relates the length of a waveguide to the maximum fabrication error allowable. In fact, assuming a maximum fabrication error of $\delta f_{max}$, from (\ref{eq:sensitivity_waveguide}) we can determine the maximum waveguide length $L_{max}$ to ensure that the process efficiency is greater than 50\%
\begin{equation}
L_{max} =  \frac{2 \Gamma}{\left|\partial_f\dbeta\right|\cdot \delta f_{max} }.
\label{eq:sensitivity}
\end{equation}

It is therefore clear that any fabrication error poses an unavoidable constraint on the waveguide length.
However, one can see that if $|\partial_f\Delta\beta|$ approaches $0$, then  $\delta f_{max}$ tends to infinity. Under this condition, the waveguide becomes first-order insensitive to the fabrication parameter variations. The condition $\partial_f\Delta\beta = 0$ is known as \textit{noncritical phasematching} and has been investigated in detail in previous works \cite{Lim1990,Bortz1994}.

Two main conclusions can be drawn from the analysis presented in this section. Firstly, for a desired process, there is an inverse proportionality between maximum waveguide length and fabrication errors. This means that the technological accuracy poses a well defined limit on the maximum length of the waveguides. The second conclusion is that technological imperfections can be mitigated if the \textit{process sensitivity} is minimized through careful waveguide design, thereby approaching noncritical phasematching \cite{Lim1990,Bortz1994}.
It is worth stressing that these conclusions are independent of the specific waveguide technology or waveguide geometry and therefore can be applied to all systems described by (\ref{eq:ideal_pm}).
While the model has been derived for a constant fabrication error, which will not usually be the case, its predictions still provide a qualitative description of device performance in the presence of fabrication imperfections, as we will show in the following sections.

\subsection{Numerical analysis of $|\partial_w\Delta\beta|$ for Ti:LN waveguides}
\label{subsec:numericalanalysis}
We now apply the previous theory to titanium in-diffused lithium niobate (Ti:LN) waveguides in order to study the technological limits of this platform.
Ti:LN waveguides have been widely used for classical and quantum applications \cite{Regener1988, Amin1997, Kanbara1999, Thyagarajan2009,Eckstein2011, Luo2015, Stefszky2017, Stefszky2018}. They exhibit extremely low propagation losses ($<$ 0.1 dB/cm), can guide both TE and TM polarization modes, possess high nonlinearity, allow on-chip manipulation of the light field via integrated beamsplitters and acousto- and electrooptical modulators and can be easily interfaced to fibre network via pigtailing \cite{Montaut2017}.

\begin{figure}[tbp]
\begin{minipage}[t]{.45\textwidth}
    \centering
    \includegraphics[width=.6\textwidth]{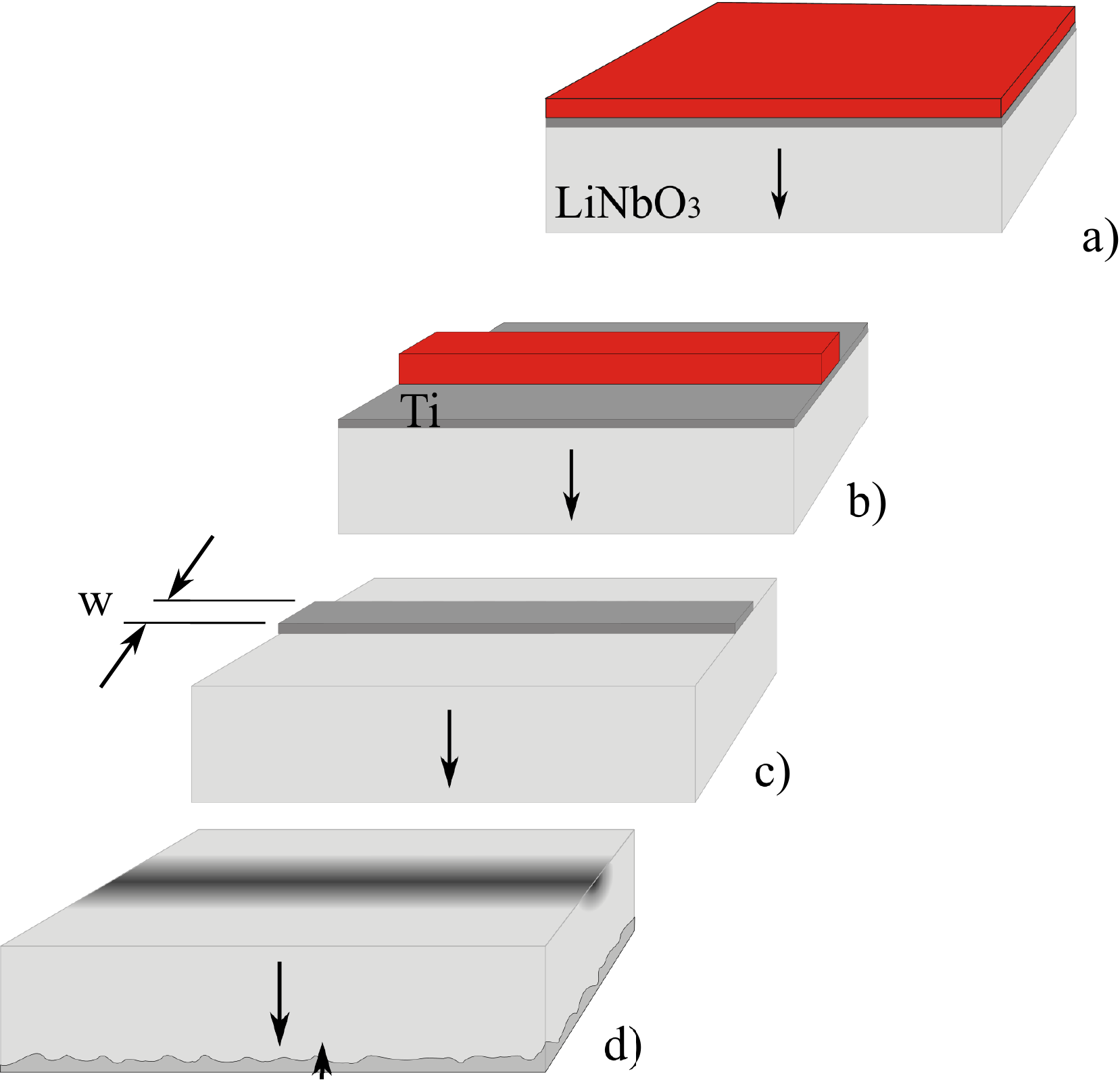}
    \caption{Standard Ti:LiNbO${}_3$ waveguide fabrication technique. From top to bottom: a) deposition of a titanium layer on top of the lithium niobate substrate and spin coating of a photoresist layer; b) photolithographic patterning of the photoresist; c) etching of the titanium to define the Ti stripe to be diffused; d) diffusion of the Ti stripe in the substrate to define the waveguide.}
    \label{img:wg_fabrication}
\end{minipage}\hfill%
\begin{minipage}[t]{.45\textwidth}    
    \centering
    \includegraphics[width=\textwidth]{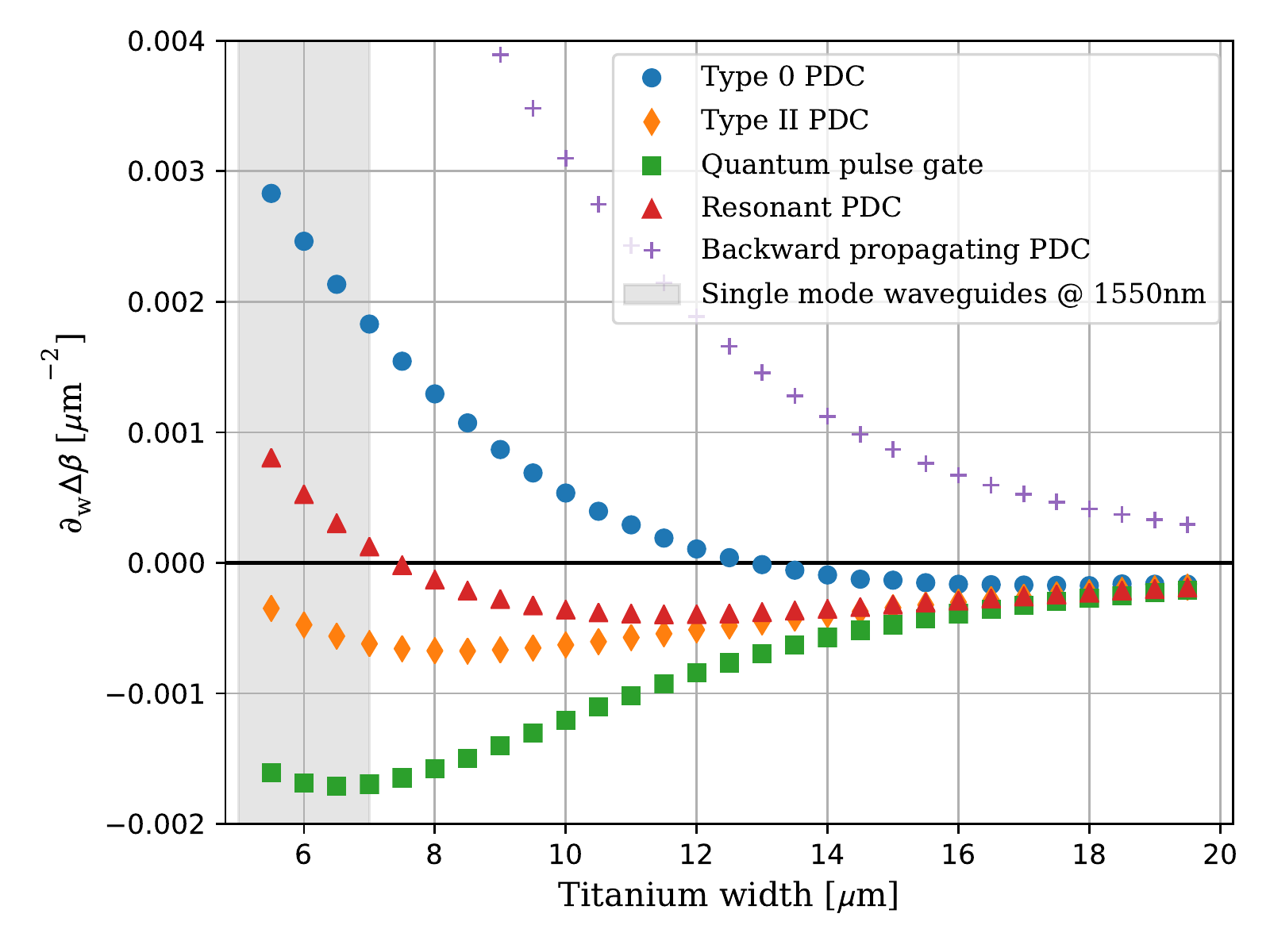}
    \caption{Calculated sensitivity $\partial_w\dbeta$ of different processes to variations of the Ti stripe width for Ti:LN waveguides. The processes analysed are: type-0 PDC (775nm$\rightarrow$1550nm, e$\rightarrow$ee), type-II PDC (775nm$\rightarrow$1550nm, o$\rightarrow$eo), quantum pulse gate \cite{Eckstein2011, Allgaier2017} (1550nm+860nm$\rightarrow$553nm, oe$\rightarrow$o), resonant PDC \cite{Luo2015} (532nm$\rightarrow$890nm+1320nm, o$\rightarrow$eo) and counter-propagating PDC (765nm$\rightarrow$1510nm+1550nm, e$\rightarrow$ee). }
	\label{img:sensitivity_LN}
\end{minipage}
\end{figure}

As illustrated in Figure \ref{img:wg_fabrication}, Ti:LN waveguides are produced by photolithographic patterning of titanium stripes with definite widths $w$ and thicknesses $\tau$ on top of a LiNbO${}_3$ substrate.
Subsequently, titanium is diffused inside the LN lattice by heating the sample in an oven. 
The resulting waveguide is defined by the initial titanium stripe geometry, the exchange temperature and the exchange time. 
Finally, periodic poling is performed by electric field poling after photolithographic patterning of the electrodes on the crystal faces. 
Fabrication errors can occur at different steps, e.g. inhomogeneous illumination conditions can affect the patterning of the titanium stripes, or temperature gradients in the diffusion oven can lead to inhomogeneous diffusion of the titanium.
All these imperfections add up and can cause local deviations of the waveguide profile with respect to the ideal, homogeneous case.

Here we use (\ref{eq:sensitivity}) to study qualitatively the fabrication limits of Ti:LN waveguides. 
To simplify the treatment, we choose to consider only one source of error, namely variation of the width $w$ of the Ti stripe (from now on we will refer to $w$ simply as the waveguide width).

In order to estimate $\partial_w\Delta\beta$, the effective refractive index of the guided modes as a function of the wavelength, polarization and waveguide width is needed. We employ a finite element solver written in Python implementing the model described in \cite{Strake1988} to calculate the Sellmeier equations of waveguides  produced with different widths $w$.
The process sensitivity $\partial_w \dbeta$ as a function of $w$ for different processes is shown in Figure \ref{img:sensitivity_LN} for a number of processes of interest for quantum applications, namely: 
type-0 PDC \cite{Stefszky2017}; type-II PDC \cite{Sansoni2017}; the quantum pulse gate (QPG) \cite{Eckstein2011}; the resonant PDC source described in \cite{Luo2015} and counter-propagating PDC generating photons at 1510nm and 1550nm.

\begin{figure}[btp]
    \centering
    \subfloat[Type-0 PDC 775nm$\rightarrow$1550nm]
    {
    \includegraphics[width=.4\textwidth]{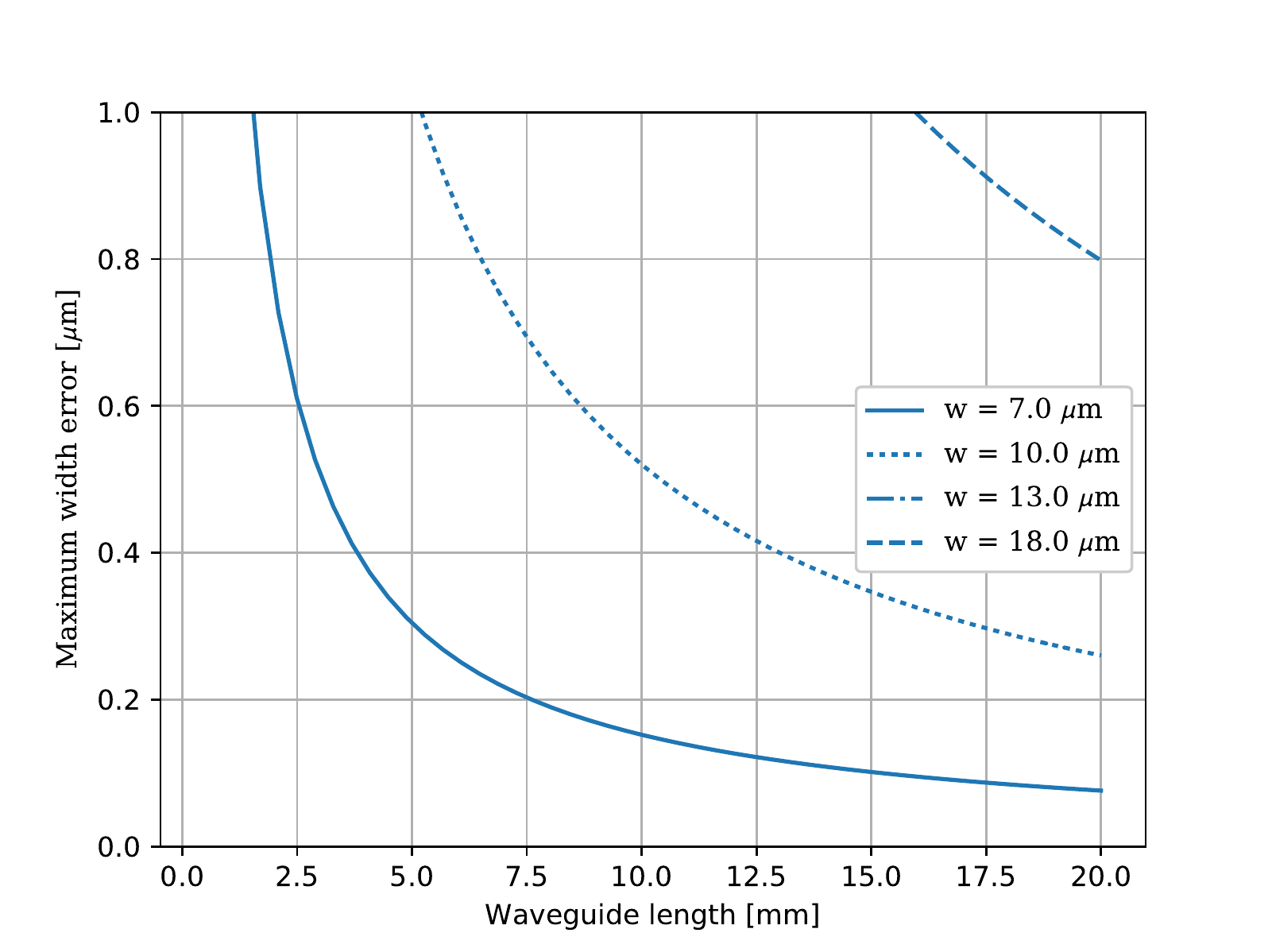}
    \label{subpl:LNSHG0}} \quad
    \subfloat[Quantum pulse gate]
    {\includegraphics[width=.4\textwidth]{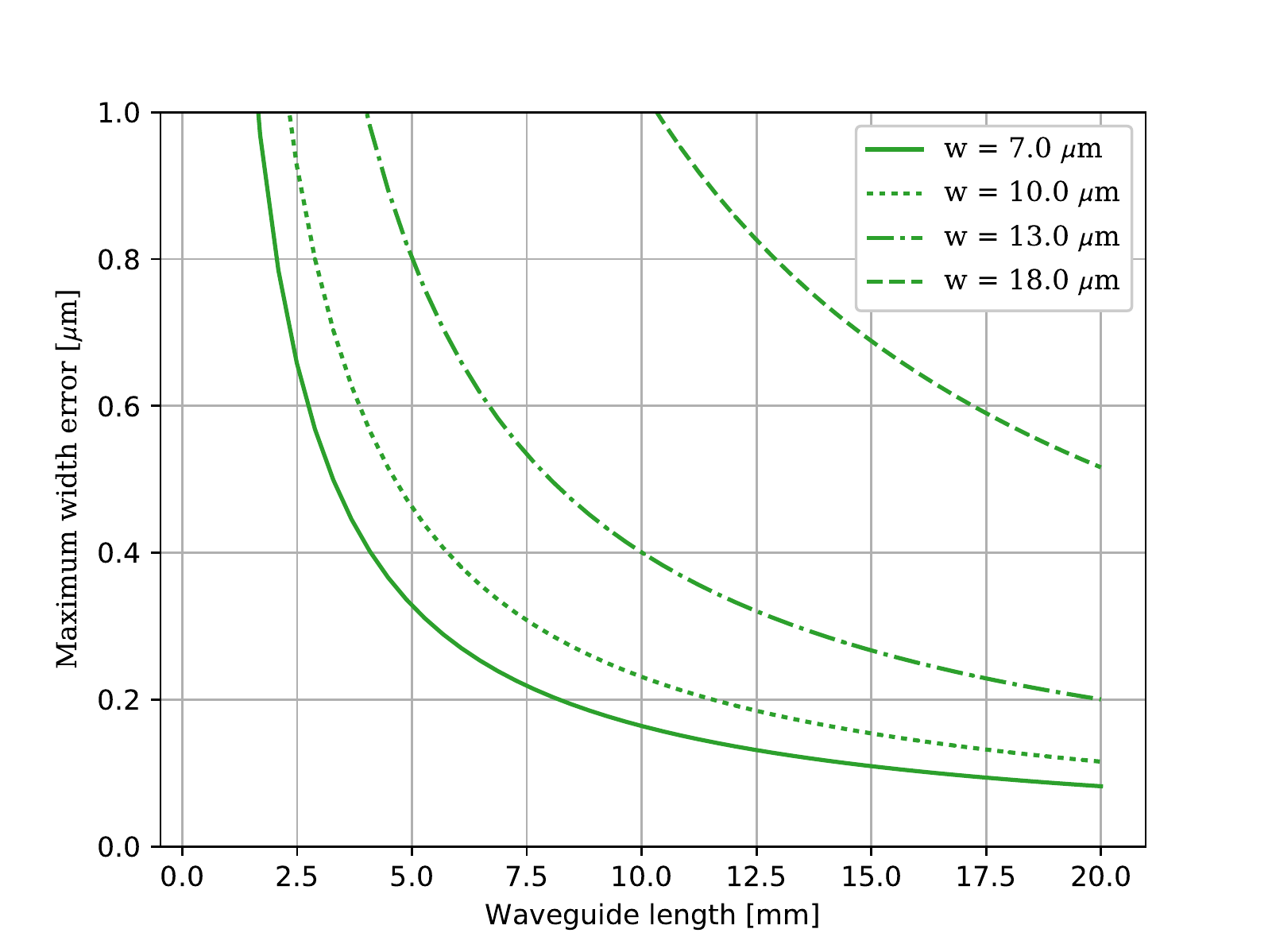}
    \label{subpl:LNQPG}}
    \caption{The dependence of the maximum width error on the chosen waveguide length is shown for type-0 PDC (\ref{subpl:LNSHG0}) and for the quantum pulse gate (\ref{subpl:LNQPG}). This dependence is shown for different waveguide widths. The plots show that the maximum allowable width error decreases as $1/L$ and that wider waveguides are less sensitive to the width error. Note that the 13$\mu$m line in Figure \ref{subpl:LNSHG0} is not present because it is first-order immune to noise.}
    \label{img:LNsensit}
\end{figure}

Recall that the waveguide is first-order immune to noise if the condition $\partial_w\Delta\beta=0$ is met. 
Among the processes considered, only the resonant PDC process is non-critically phasematched in a regime where the waveguide is single-mode at telecom wavelengths. The type-0 PDC process is noncritically phasematched for $w$=13$\mu$m, but the waveguide is spatially multimode for this width. This is unfortunate as single-mode operation is often required.

Another important observation is that each process has a different sensitivity; even the ones involving similar wavelengths exhibit very different behaviour, e.g.  type-0 and type-II PDC. Therefore, the process sensitivity has to be investigated independently for every process under consideration. 

Using the calculated process sensitivities, we can estimate the maximum tolerable width error depending on the desired sample length using (\ref{eq:sensitivity}).
The results for type-0 PDC and the QPG are displayed in Figure \ref{img:LNsensit}. The model predicts that width errors of $|\delta w|\leq 0.2 \mu$m already limit the maximum waveguide lengths for these two processes to around $10$mm. The results for the other processes are reported in the supplementary material.
As we will show in the following section, this simple estimation still provides a good indication of the region where fabrication imperfection may start to play a role. 

\begin{figure}[tbp]
    \centering
    \includegraphics[width=0.5\textwidth]{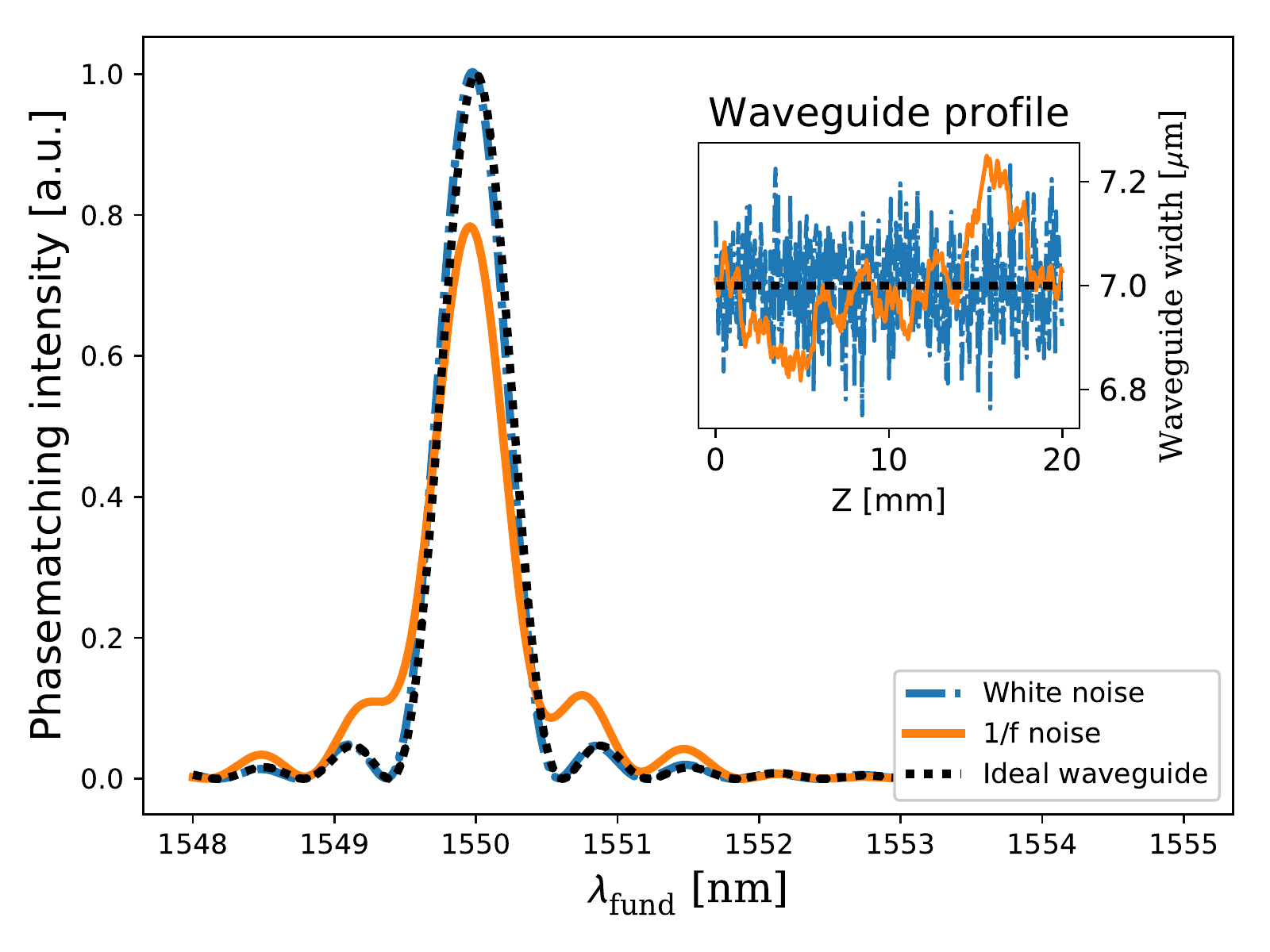}
    \caption{Example of the effect of different noise profiles on the phasematching intensity spectrum. In the main plot, the phasematchings of the waveguides without noise (black dotted line), with additive white gaussian noise (blue dashdotted line) and with $1/f$ noise spectrum (orange solid line) are shown. In the inset, the respective waveguide width profiles are reported. The device under consideration is a 20mm-long Ti:LiNbO${}_3$ waveguide for type-0 PDC 775nm $\rightarrow$ 1550nm, characterized via the reverse process, SHG.}
    \label{img:noise_comparison}
\end{figure}


\section{Phasematching in inhomogeneous guiding structures.}
\label{sec:inhomogeneousWaveguides}
\subsection{Impact of different noise profiles on the phasematching.}
\label{subsec:statanalysis}
The analysis conducted in the previous sections considered only homogeneous waveguides. 
However, in reality fabrication errors can occur randomly along the waveguide, thus leading to the production of \textit{inhomogeneous} waveguides, where the refractive index distribution varies along $z$. 
For this reason, a spatially-varying fabrication parameter $f_{prod}(z)$ leads to a momentum mismatch $\Delta\beta(z)$ that varies along the waveguide. 
In this case equation (\ref{eq:ideal_pm}) does not hold anymore and a more general expression has to be considered \cite{Helmfrid1992}
\begin{equation}
\phi \propto \frac{1}{L}\int_0^L \rme^{\rmi\int_0^z\Delta\beta(\xi) \rmd\xi} \rmd z.
\label{eq:general_pm} 
\end{equation}
Integration of (\ref{eq:general_pm}) is possible usually only numerically and by assuming specific profiles for the momentum mismatch variation $\Delta\beta(z)$ along the waveguide. 
Moreover, the phasematching spectrum will not result in the usual $\mathrm{sinc}^2$ shape \cite{Helmfrid1991,Pelc2010,Pelc2011a,Phillips2013}. 

In the past, investigation of waveguides with variable dispersion profiles has been restricted to classical SHG systems assuming simple profiles for $\Delta\beta(z)$ \cite{Helmfrid1991}.
On the other hand, random fabrication errors may dramatically affect the desired quantum state produced in waveguide systems. 
Therefore, in the remaining sections we study the effect of randomly variable dispersion relations in waveguides designed for quantum processes. 

Here we study the phasematching properties of inhomogeneous Ti:LN waveguides as a function of the Ti stripe width $w$ and its maximum error $\delta w$. 
Generating different profiles for $w(z)$ and calculating the relative momentum mismatch $\Delta\beta(z)$, we can integrate numerically (\ref{eq:general_pm}) to calculate the relative phasematching spectra. 
The details of the simulations of this section are presented in the supplementary material, section 2.

For simplicity, we investigate fabrication errors $\delta w(z)$ characterized by two types of noise spectra, namely additive white gaussian (AWG) and $1/f$ noise. AWG noise describes uncorrelated noise fluctuations along the waveguide, while $1/f$ noise is characterised by spatial correlations and accounts for long-range drifts in the production parameters. 
An example of how these noise spectra affect the width profile and the phasematching spectra is shown in Figure \ref{img:noise_comparison}.

\begin{figure}[tbp]
  \centering
  	\includegraphics[width=0.5\textwidth]{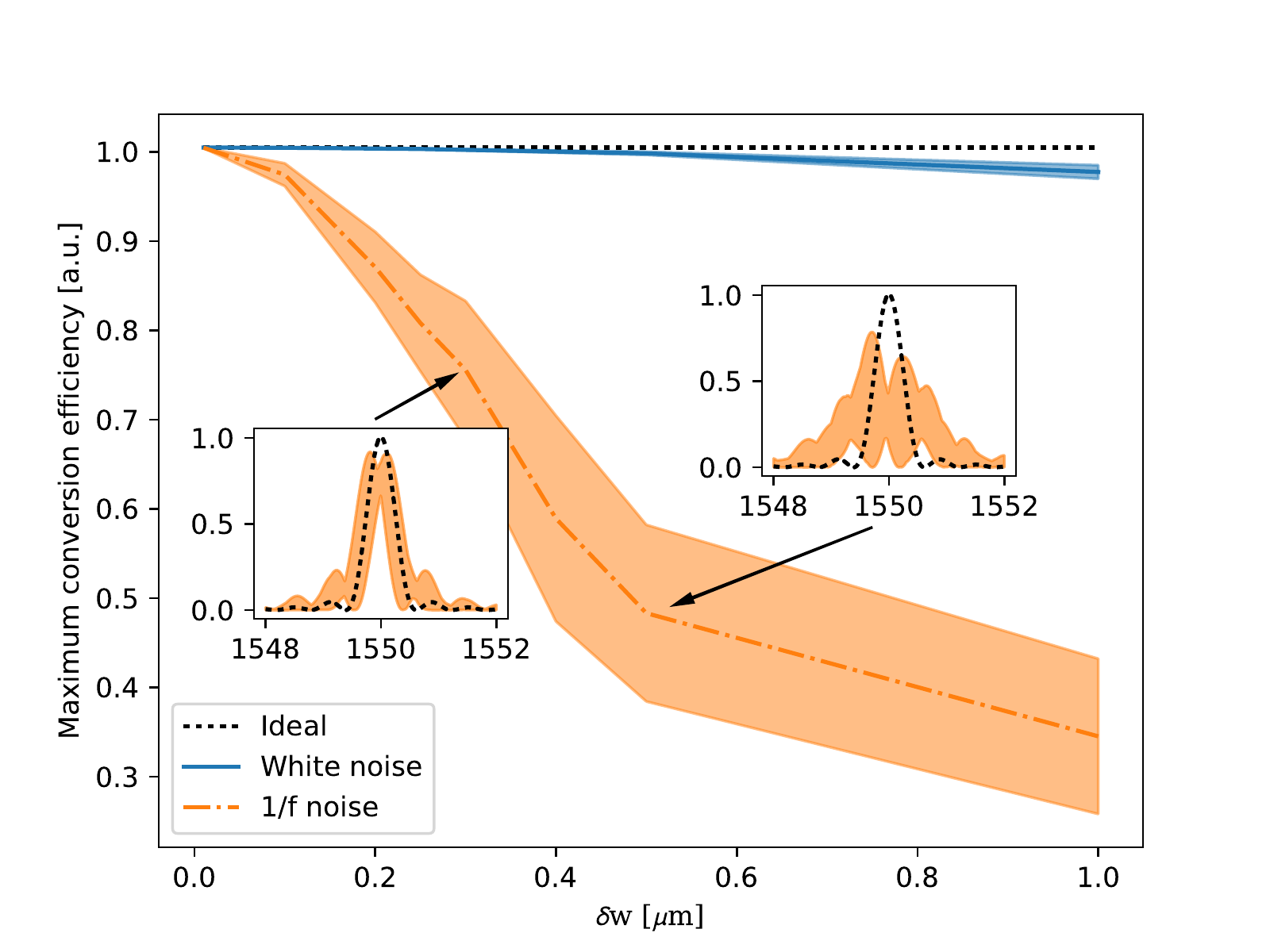}
   	\caption{Maximum efficiency of a nonlinear process as a function of the waveguide width error for the ideal waveguide (black dotted line), a waveguide with AWG noise (blue dashdotted line) and $1/f$ noise (orange solid line). Shaded regions correspond to the range of results of 40 simulations for each datapoint. 
   	The insets show the effect of $1/f$ noise on the phasematching intensity in comparison to an ideal waveguide.
	The shaded area represent the range of simulated intensity spectra from 40 simulations.
	A broadening of the average phasematching spectrum, more prominent side lobes and reduction of the efficiency are evident.  
   	The device under consideration is a 20mm-long Ti:LiNbO${}_3$ waveguide for type-0 PDC 775nm $\rightarrow$ 1550nm, characterized via the reverse process, SHG.}
    \label{img:comparison_awgn_pink}
\end{figure}

To understand the main differences between the two types of noise, we study the performance of a 20mm-long, 7$\mu$m-wide waveguide designed for a type-0 second harmonic generation (SHG) pumped at 1550nm. We investigate the degradation of the conversion efficiency for values of the fabrication error $\delta w \in [0, 1.0]\mu$m for both types of noise.
The results of the simulations are presented in Figure \ref{img:comparison_awgn_pink}.
The two types of noise have very different impact on the maximum achievable conversion efficiency: AWG noise has a negligible influence, while $1/f$ noise can drastically decrease it. 
Furthermore, the reduction of conversion efficiency is accompanied by an increase of the phasematching bandwidth, especially for errors $\delta w>0.25 \mu$m, whose broadened phasematching is shown in the insets of Figure \ref{img:comparison_awgn_pink}.

The same analysis has been performed for the other processes characterized in Figure \ref{img:sensitivity_LN} and the results are similar: AWG noise consistently has a negligible impact on the average maximum conversion efficiency, while $1/f$ noise rapidly degrades the performance of the device as $\delta w$ increases.
These results are well in agreement with previous studies on different systems. The presence of AWG noise on the poling grating of periodically-poled waveguides has been previously analysed in \cite{Pelc2010, Pelc2011a, Phillips2013} and showed only a minor influence on the maximum conversion efficiency. Moreover, a comparison between correlated and uncorrelated noise has been investigated in photonics crystal fibres, showing that imperfections with long-range correlations drastically effect the parametric gain of nonlinear processes \cite{Farahmand2004}.
In the rest of the paper, we will focus our attention exclusively on the effects of $1/f$ noise, since this is the main cause of phasematching distortions.

\begin{figure}[tbp]
    \centering
    \includegraphics[width=0.5\textwidth]{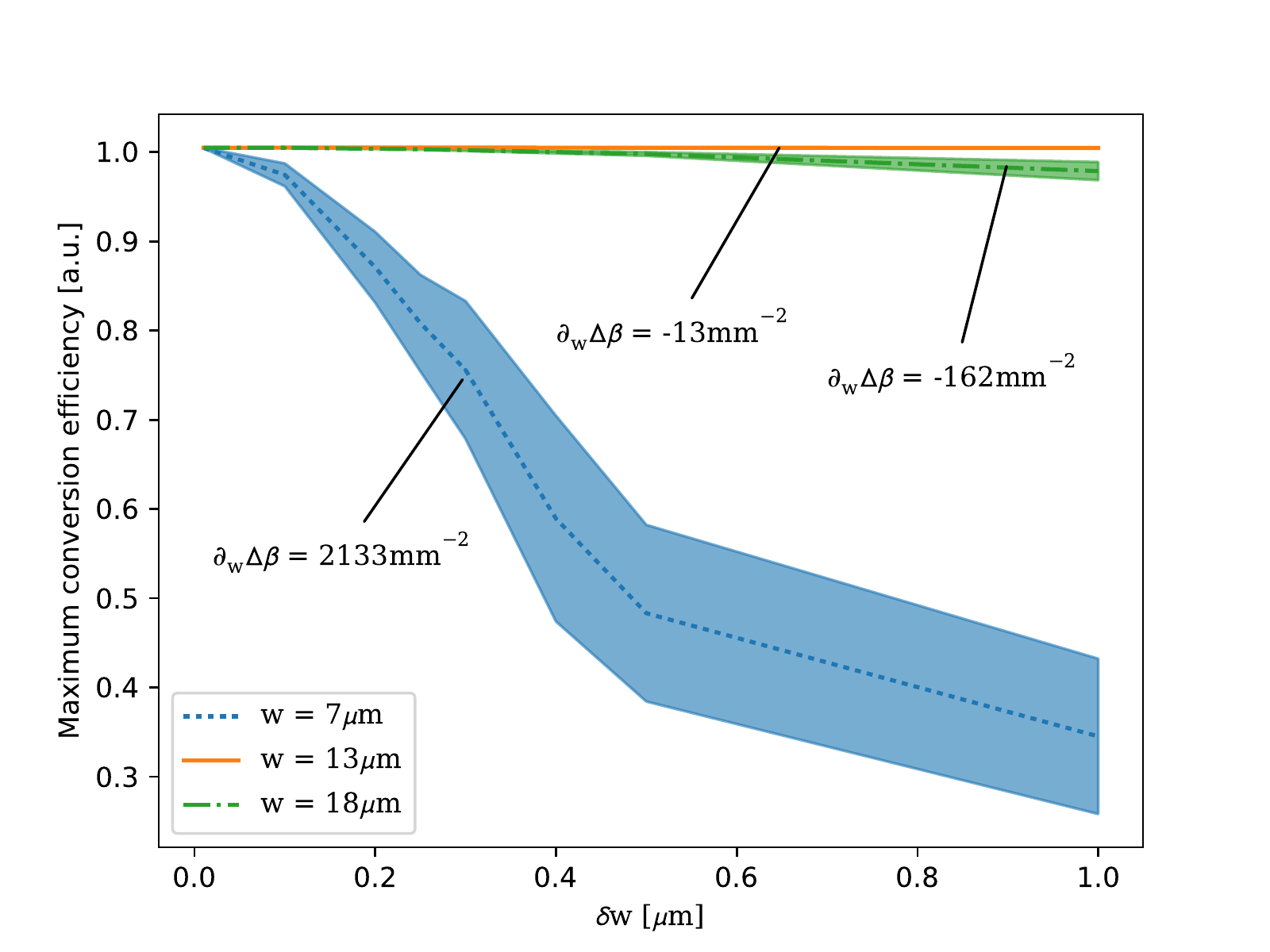}
    \caption{Maximum efficiency of a nonlinear process as a function of the waveguide width error, in the presence of $1/f$ noise. The three different curves are calculated for nominal widths of 7$\mu$m (blue dotted line), 13$\mu$m (orange solid line) and 18$\mu$m (green dashed line). Shaded regions are the errorbars as retrieved from stochastic simulations. Note that the 13$\mu$m waveguide is noncritically phase matched and so is virtually immune to the presence of noise. The device under consideration is a 20mm-long Ti:LiNbO${}_3$ waveguide for type-0 PDC 775nm $\rightarrow$ 1550nm, characterized via the reverse process, SHG.}
    \label{img:SHG0_stat_analysis}
\end{figure}

Having established a framework suitable for the study of waveguide inhomogeneities, we can now compare the approximate results derived in section \ref{sec:qualitativeModel} with numerical simulations.
An important result was the ability to predict design of noncritically phasematched waveguides. In particular, we calculated that a type-0 PDC process pumped at 775nm is noncritically phasematched for $w=13\mu$m.
This result is confirmed by evaluating the conversion efficiency of the reverse process, a type-0 SHG pumped at 1550nm, as a function of the waveguide width $w$ and the error $\delta w$ in presence of $1/f$ noise. 
Indeed, Figure \ref{img:SHG0_stat_analysis} shows that a 13$\mu$m-wide waveguide is practically immune to $1/f$ noise when close to noncritical phasematching.
Moreover, from the calculations reported in Figure \ref{subpl:LNSHG0}, we expect that a 7$\mu$m-wide, 20mm-long waveguide will be sensitive to noise values $\delta w\geq 0.1 \mu$m. As shown in Figure \ref{img:SHG0_stat_analysis}, for $\delta w \geq 0.1\mu$m, the maximum efficiency rapidly degrades below 90\% of the ideal maximum. This confirms that the simplified model can provide reliable qualitative information about the waveguides' sensitivity to noise and thus the evaluation of the process sensitivity is can provide useful technological boundaries for the process quality of waveguide production.

\subsection{Applications}
\label{subsec:applications}
The theory presented so far is now applied to three different systems of interest in quantum optics.  In fact, we will show that it is necessary to consider the impact of fabrication errors in these systems to correctly model and estimate their performance.
In section \ref{subsec:squeezing} we analyse the effect of waveguide inhomogeneities on the maximum squeezing attainable in a waveguided system; in section \ref{subsec:hdqkd} we estimate how noise reduces the maximum number of bins in a frequency-bin encoding (FBE) scheme; in section \ref{subsec:QPG} we study the effect of waveguide width noise on the bandwidth compression factor of a frequency conversion device.

\subsubsection{Impact of fabrication errors in squeezing generation.}
\label{subsec:squeezing}
We first consider a waveguide structure designed to produce continuous-wave (CW) single-mode squeezed states in a single-pass configuration. 
These states are the foundation for continuous-variable (CV) quantum optics: they can be used as a basis for CV quantum computing \cite{Menicucci2006}, they have been used to generate complex quantum states such as EPR entanglement \cite{Bowen2002} and CV cluster states \cite{Yukawa2008}, and they have been used in sensing and metrology in order to improve the sensitivity of measurements, e.g. in gravitational-wave astronomy \cite{Aasi2013}.

We consider here  a 7$\mu$m-wide Ti:LN waveguide pumped at 775nm that produces type-0 squeezing at 1550nm in a single-pass configuration. 
It can be shown that both the losses of the fundamental field and the strength of the nonlinear process are critical to the amount of squeezing produced \cite{Serkland1995}. 
We begin by neglecting the losses, thereby exclusively investigating the effect of waveguide width imperfections on the strength of the nonlinear process. 
The strength of the nonlinear process can be found by performing second harmonic generation in such a sample, from which one can calculate the normalized conversion efficiency using
\begin{equation}
\eta_{norm} =\frac{P_{SH}}{P_{FF}^2 L^2}.
\label{eq:normalized_conversion_efficiency}
\end{equation}
A common misconception is that, due to its definition, $\eta_{norm}$ is independent of length.
However, the qualitative model presented in Figure \ref{subpl:LNSHG0} shows that longer waveguides are more susceptible to fabrication imperfections, therefore we expect $\eta_{norm}$ to be dependent on waveguide length.

\begin{figure}[tbp]
\begin{minipage}[t]{.45\textwidth}
    \centering
    \includegraphics[width=1.\textwidth]{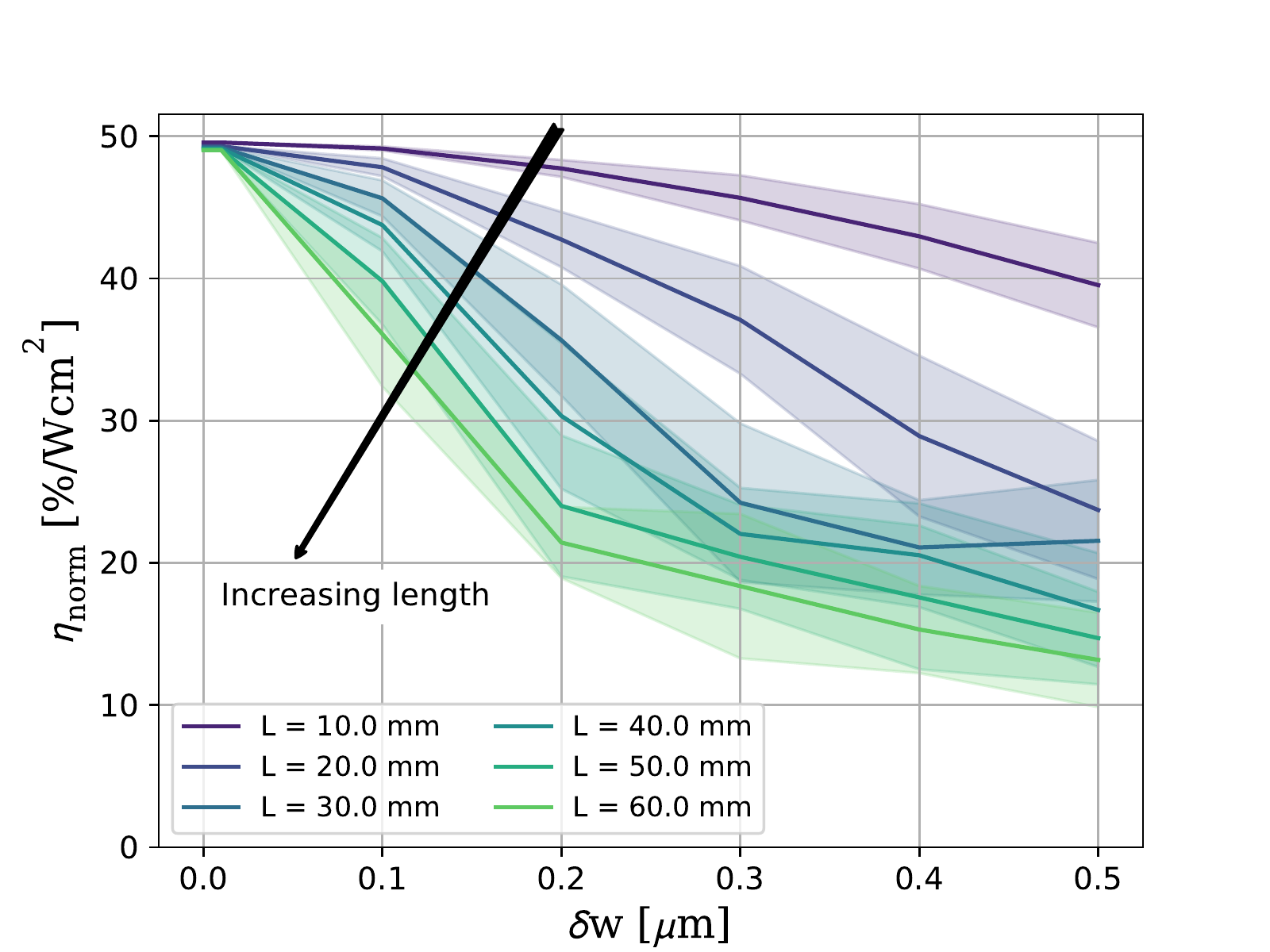}    
    \caption{Normalised conversion efficiency as a function of the error on the waveguide width, for lossless samples of varying lengths. It can be seen that both the length of the sample and the magnitude of the width error have a strong impact on the normalized conversion efficiency, even in the absence of losses. The device under consideration is a 7$\mu$m-wide  Ti:LiNbO${}_3$ waveguide for type-0 PDC 775nm $\rightarrow$ 1550nm, characterized via the reverse process, SHG.}
        \label{subimg:norm_conv_eff_shg0}
\end{minipage}\hfill%
\begin{minipage}[t]{.45\textwidth}
	\centering
    \includegraphics[width=1.\textwidth]{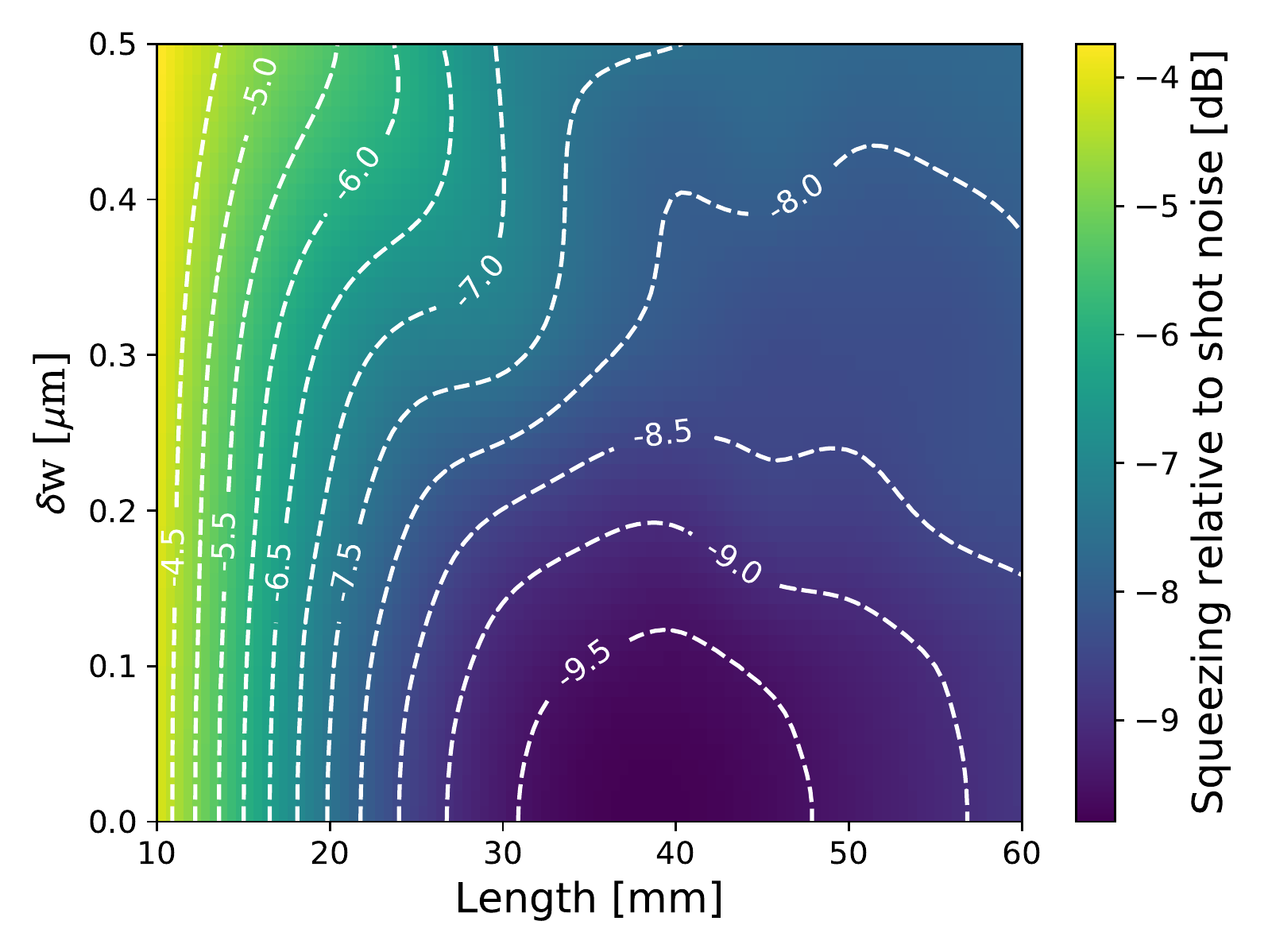}    
    \caption{Squeezing relative to shot noise exiting the waveguide for samples with different lengths and widths error. The process is pumped with 500mW of CW input at 775nm. Losses for the fundamental field are assumed to be 0.1 dB/cm.  The device under consideration is a 7$\mu$m-wide Ti:LiNbO${}_3$ waveguide for type-0 PDC 775nm $\rightarrow$ 1550nm}
    \label{subimg:squeezing}
\end{minipage}
\end{figure}

To calculate $\eta_{norm}$ in presence of fabrication errors, we numerically simulate the phasematching spectra of the system for different sample lengths $L\in$ [10, 60]mm and width error magnitude $\delta w\in$ [0, 0.5]$\mu$m. 
For each parameter combination, we calculate the maximum conversion efficiency of 40 randomly generated systems to estimate the average normalized conversion efficiency.
From Figure \ref{subimg:norm_conv_eff_shg0}, it is evident that the normalized conversion efficiency is critically dependent on both $L$ and $\delta w$.
The simulations reveal that both $\eta_{norm}$ and the waveguide length of each sample are necessary to fairly compare the performance of different devices.
Furthermore it can be seen that the normalized conversion efficiency drops from 49 \%/Wcm$^2$ to 40\%/Wcm$^2$ for 10mm-long waveguides and below 15\%/Wcm$^2$ for 60mm-long waveguides.
Therefore, one will overestimate the benefits of longer samples if the impact of fabrication errors is not taken into account.

From the calculated normalized conversion efficiencies, one can estimate the amount of squeezing that can be produced in this device. 
Follwing \cite{Serkland1995}, the maximum squeezing $S$ achievable in a single-pass CW waveguide can be given by
\begin{equation}
S=\left(\rme^{-2\sqrt{\eta_{norm}P_{in}}L} \rme^{-\alpha L}\right)+1-\rme^{-\alpha L},
\end{equation}
where $P_{in}$ is the input pump power of the squeezer and $\alpha$ is the loss for the squeezed field. We assume a negligible effect of the losses for the 775nm pump.
We consider $P_{in}$ = 500mW at 775nm and propagation losses $\alpha$ equal to 0.1 dB/cm, a safe estimate of the average losses measured in Ti:LN waveguides \cite{Stefszky2017}. 
The squeezing $S$ produced as a function of $L$ and $\delta w$ is shown in Figure \ref{subimg:squeezing}. 
It can be seen that, for a given $\delta w$, there exists a waveguide length that maximizes the squeezing produced.
Moreover, this optimal length increases as the magnitude of the width error increases.
This is due to a complex interplay between the nonlinear interaction strength and the losses; as the waveguide length increases, the positive effect of an increase in the interaction length is counteracted by an increase in the total losses and a simultaneous reduction of the normalized conversion efficiency.
The simulations show that the system under investigation (with 500mW of pump power) can produce around -9.5 dB of squeezing, choosing a waveguide with an optimized length of 40mm, if the error is below $\delta w\leq$ 0.1$\mu$m.

To reduce the impact of fabrication errors, one can consider the use of noncritically phasematched systems. 
For the system under consideration, this can be done by choosing a 13$\mu$m-wide waveguide, as shown in Figures \ref{img:LNsensit} and \ref{img:SHG0_stat_analysis}.
In this case we expect a normalized conversion efficiency that is independent of the waveguide length and the fabrication imperfections. 
Note that insensitivity to fabrication imperfections is equivalent to having no fabrication imperfection.
Therefore, the squeezing produced in a noncritically phasematched waveguide corresponds to the values at $\delta w =0\mu$m in Figure \ref{subimg:squeezing}, neglecting a minor deviation in $\eta_{norm}$ due to differences in the overlap of the interacting fields in the wider waveguide.

\subsubsection{Impact of fabrication errors on quantum information encoding.}
\label{subsec:hdqkd}
Frequency-bin encoding (FBE) is an attractive scheme for implementation of quantum information processing protocols because it offers an unbounded space for high-dimensional encoding compatible with standard fibre networks. 
Furthermore, FBE can be implemented using PDC sources, which are a versatile and tunable platform that has been developed for many years.

Here, we study the limitations of Ti:LN waveguides as pulsed PDC sources for FBE and evaluate the impact of fabrication imperfections on such systems. 
We consider a type-0 PDC source in a Ti:LN waveguide, pumped with at 775nm, generating pairs of frequency-bin entangled photons in the telecom C-band, between 1530nm and 1570nm. 
The physical device is analoguous to the one presented in \cite{Olislager2010}.
Typically, PDC sources for FBE are pumped with CW light \cite{Olislager2010,Zhong2015}, however, pulsed systems provide advantages in terms of synchronization between the communicating parties.
Therefore, we consider a pulsed pump laser with a pump bandwidth broader than the phasematching.

\begin{figure}[tbp]
\begin{minipage}[t]{.45\textwidth}
    \centering
    \includegraphics[width=1.\textwidth]{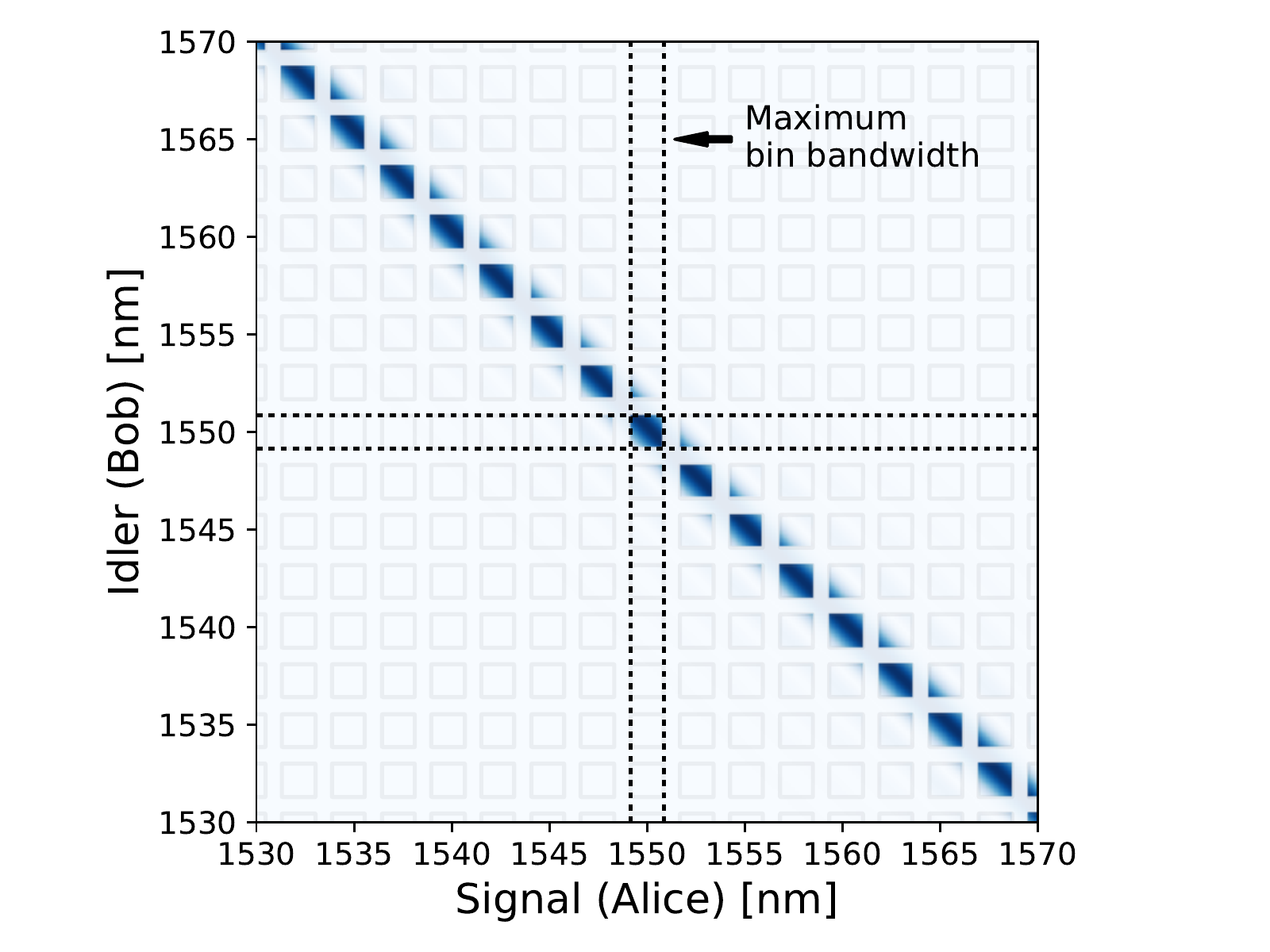}
    \caption{Definition of maximum bin bandwidth. A simple criterion to ensure low cross-talk between the different frequency bins is to define the bin size equal to the FWHM of the phasematching bandwidth, and distance between bins equal to half the FWHM.}
	\label{img:bin_bandwidth}
\end{minipage}\hfill%
\begin{minipage}[t]{.45\textwidth}
    \centering
    \includegraphics[width=1.\textwidth]{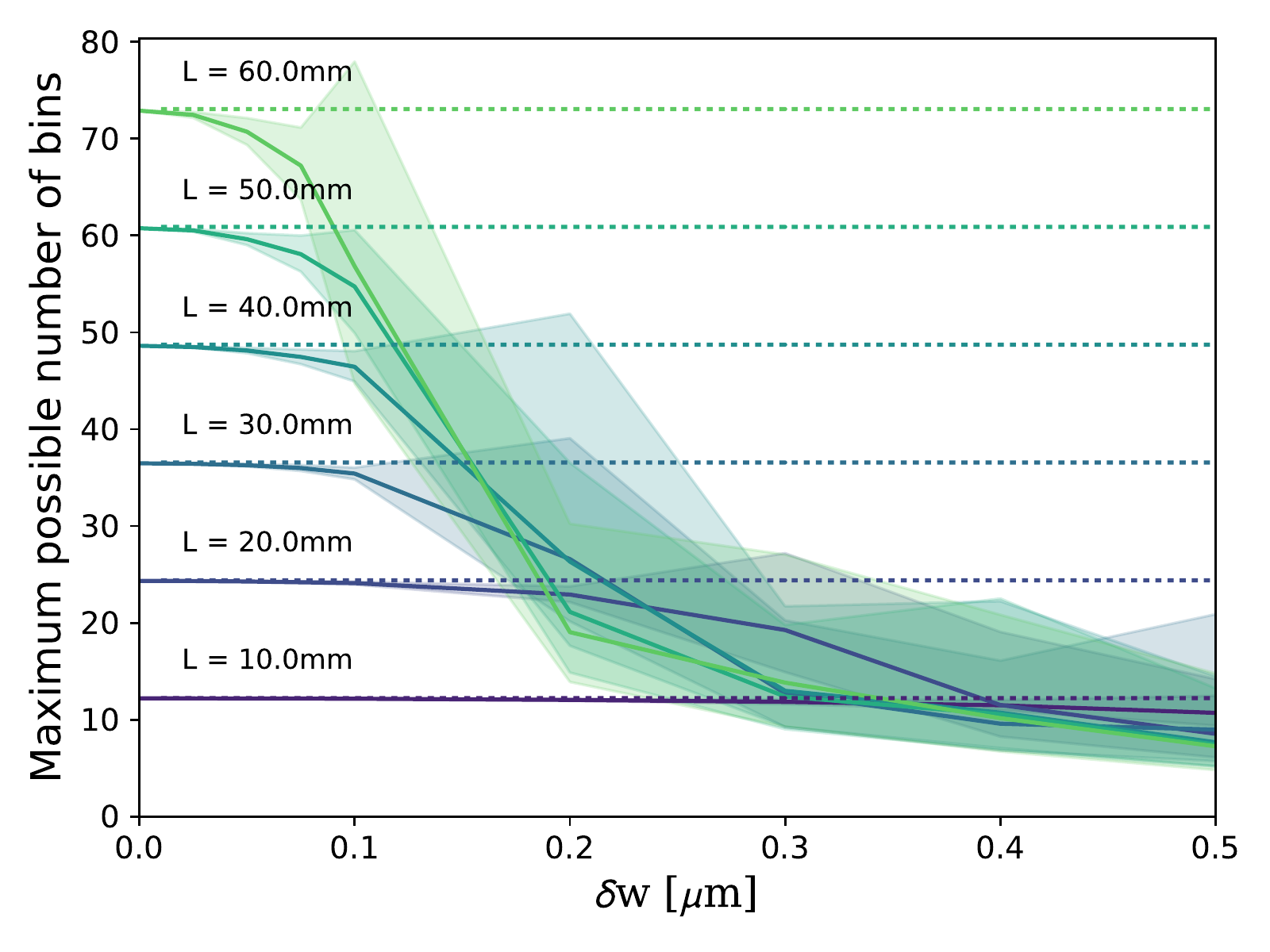}
    \caption{Maximum number of bins available for a frequency-encoded HDQKD protocol. A $1/f$ noise spectrum with maximum error $\delta w$ has been considered. 
    The bin size is set to be equal to the phasematching FWHM and the distance between the channels is half the FWHM. Solid lines are for a 7$\mu$m-wide waveguide, dashed line for a 13$\mu$m-wide waveguide.}
	\label{img:numb_bins}
\end{minipage}
\end{figure}

FBE benefits from having a large number of encoding bins; however, it is also important to minimize cross-talk between them. 
As a compromise between these two factors, we define the frequency-bin bandwidth $\Delta b$ as the full width at half maximum (FWHM) of the phasematching spectrum and each bin is separated by $\Delta b/2$, as illustrated in Figure \ref{img:bin_bandwidth}.
From this definition, given the available frequency band $\Delta\lambda$, one can calculate the number of available bins $n_{bins}$
\begin{equation}
n_{bins} = \frac{\Delta\lambda}{1.5\Delta b},
\end{equation}
where $\Delta\lambda=40nm$ is the bandwidth of the telecom C-band.
The number of available bins is then used as a figure of merit for the system. 
The bin bandwidth $\Delta b$ is extracted from the phasematching spectrum by fitting it with a Gaussian and taking the FWHM of the fit.

This analysis is applied for varying sample lengths $L$ and width error magnitudes $\delta w$ and the results of the calculations are shown in Figure \ref{img:numb_bins}.
Solid lines represent 7$\mu$m-wide waveguides, while dashed ones represent 13$\mu$m-wide waveguides. Shaded areas represent the range of simulation results over 40 iterations for each data point. 
The simulations show that it is possible to implement more than 70 bins in a 60mm-long, provided that fabrication errors are minimal.
However, for the 7$\mu$m-wide waveguides, the number of bins available in longer waveguides drops rapidly with increasing width error to a minimum of approximately 10. 
In fact, for errors above 0.4$\mu$m, 10mm-long waveguides may outperform 60mm-long ones.
The reason is that longer samples theoretically have much narrower phasematching bandwidth and these are much more susceptible to fabrication errors, as highlighted in section \ref{subsec:mathdescription}.
For this reason shorter samples can outperform longer ones, in the presence of large fabrication errors.
In contrast, the 13$\mu$m-wide waveguides do not show a reduction in the number of bins as the noise increases, as can be seen from the dashed lines in Figure \ref{img:numb_bins}. 
This is due to the fact that they are noncritically phasematched and therefore immune to fabrication errors. 

\subsubsection{Impact of fabrication errors on the performance of a bandwidth compressor.}
\label{subsec:QPG}
Interfacing components operating at different wavelengths is a critical challenge for quantum optical networks.
Reduction of transmission losses is paramount in most applications and therefore transmission in the telecom C-band is desired, where losses are minimal. 
However, many quantum devices operate outside this frequency band and therefore efficient frequency conversion between these bands is required.
Furthermore, it is often necessary to efficiently match the bandwidth of different quantum devices.
Both bandwidth matching and frequency conversion can be efficiently achieved in the integrated quantum pulse gate (QPG), a device that implements type-II sum frequency generation (SFG) in a waveguide \cite{Allgaier2017}. 

The integrated QPG in \cite{Allgaier2017} was implemented in a 7$\mu$m-wide, 27mm-long Ti:LN waveguide, designed to convert single photons from the telecom C-band to 550nm. 
The measured bandwidth compression factor (BCF) was $\Delta\nu_{in} / \Delta\nu_{out} = 7.47\pm 0.01$, where $\Delta\nu_{in/out}$ is the spectral bandwidth of the input and output photons. 
In this device, the compression factor is directly related to the phasematching bandwidth: the narrower the bandwidth, the higher the compression factor. 
We have already shown in section \ref{subsec:hdqkd} that fabrication imperfections can increase the phasematching bandwidth of a given process.
Therefore, we expect that the compression factor will reduce in the presence of fabrication imperfections.

We consider a 7$\mu$m-wide waveguide with different lengths $L$ and varying magnitude $\delta w$ of $1/f$ noise on the waveguide width. 
The input bandwidth is set to  $\Delta\nu_{in} = 963\pm 11$ GHz, while the output bandwidth $\Delta \nu_{out}$ is defined as the FWHM of a Gaussian fit to the phasematching spectrum, following the method of \cite{Allgaier2017}. 

\begin{figure}[hbtp]
	\centering
	\subfloat[Bandwidth compression for a 7$\mu$m-wide waveguide.]{
    \includegraphics[width=.4\textwidth]{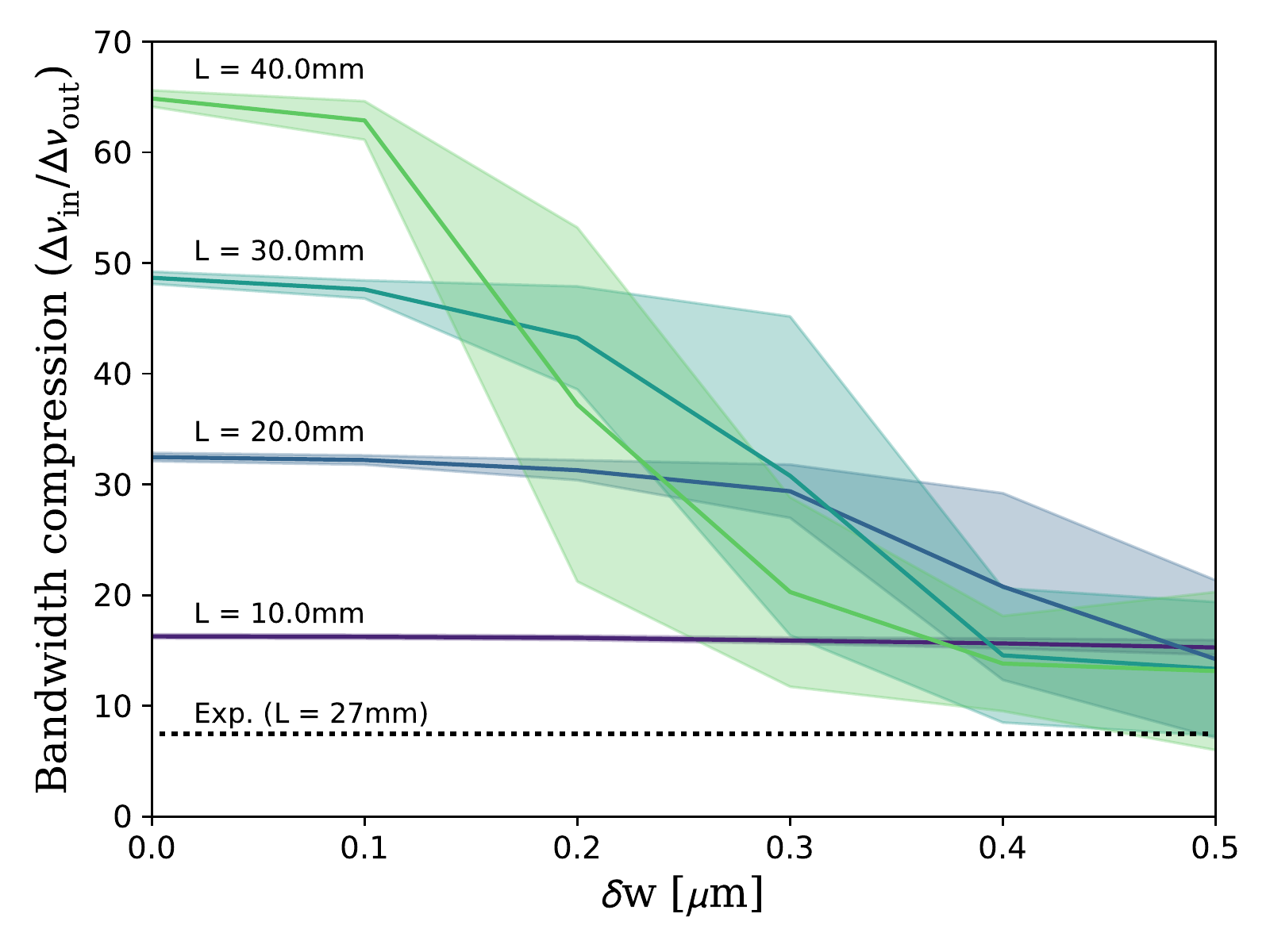}
    \label{subimg:bw_compr_7um}}\quad
    \subfloat[Bandwidth compression for a 13$\mu$m-wide waveguide.]{
    \includegraphics[width=.4\textwidth]{figure12a}    
    \label{subimg:bw_compr_13um}}
    \caption{Bandwidth compression factor resulting from the up-conversion of 963 GHz-broad telecom photons, using the quantum pulse gate \cite{Allgaier2017}. The compression factor has been calculated for different waveguide width errors $\delta w$ and lengths $L$, in presence of noise with $1/f$ noise spectrum. Moreover, two different nominal width have been investigated, namely $w=7$ $\mu$m in Figure \Ref{subimg:bw_compr_7um} and $13$ $\mu$m in Figure \Ref{subimg:bw_compr_13um}. The experimental value reported in \cite{Allgaier2017} is also shown as a dashed black line in both plots for comparison.}
	\label{img:bwcompression}
\end{figure}

Each datapoint has been simulated 40 times and the results are shown in Figure \ref{subimg:bw_compr_7um}. 
The calculated BCFs are represented in solid lines, while the shaded regions represent the range of variation in the simulated data.
Simulations show that a 40mm-long sample provides a BCF of $\sim$64, in the absence of fabrication imperfections.
This corresponds to an output bandwidth of $\sim$15GHz or, equivalently, a 30ps-long pulse, under the approximation of Gaussian  phasematching.

In Figure \ref{subimg:bw_compr_7um} it is also shown, with a dashed line, the BCF measured in \cite{Allgaier2017} for their 27mm-long waveguide. 
It is immediately evident that the measured compression factor is well below the theoretically predicted value. 
In fact, calculations show that a 27mm-long sample should provide a compression factor close to 45 in the absence of imperfections; however, the experiment measured a compression factor of only 7.45.
Such a reduction would only be expected in the presence of width error $\delta w\geq 0.4\mu$m, as illustrated in the Figure.

\begin{figure}[hbtp]
	\centering
	\subfloat[Ideal phasematching]{
    \includegraphics[width=.3\textwidth]{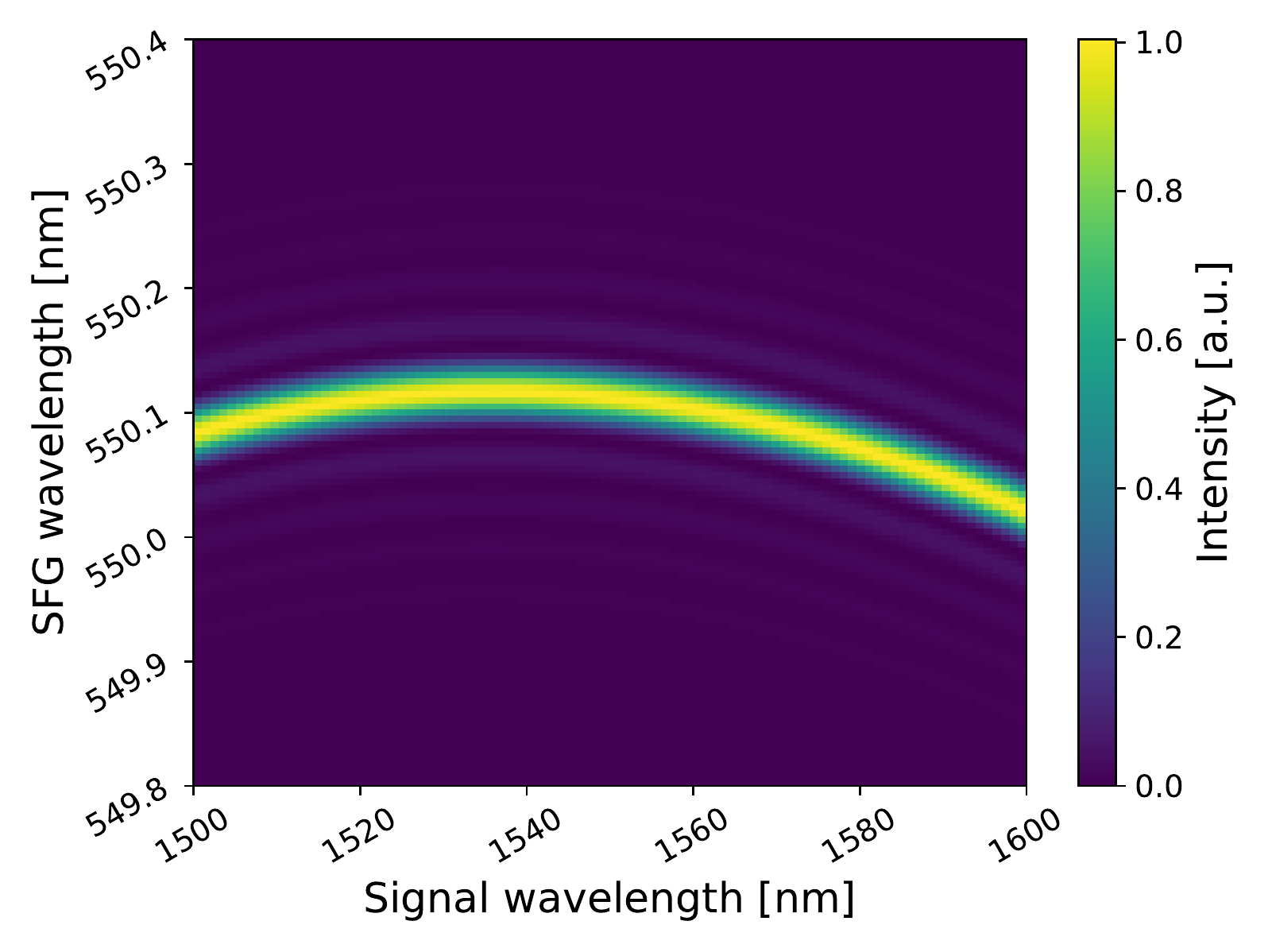}
    \label{subimg:noiseless_qpg}}\quad
	\subfloat[Phasematching measured in \cite{Allgaier2017}.]{
    \includegraphics[width=.27\textwidth]{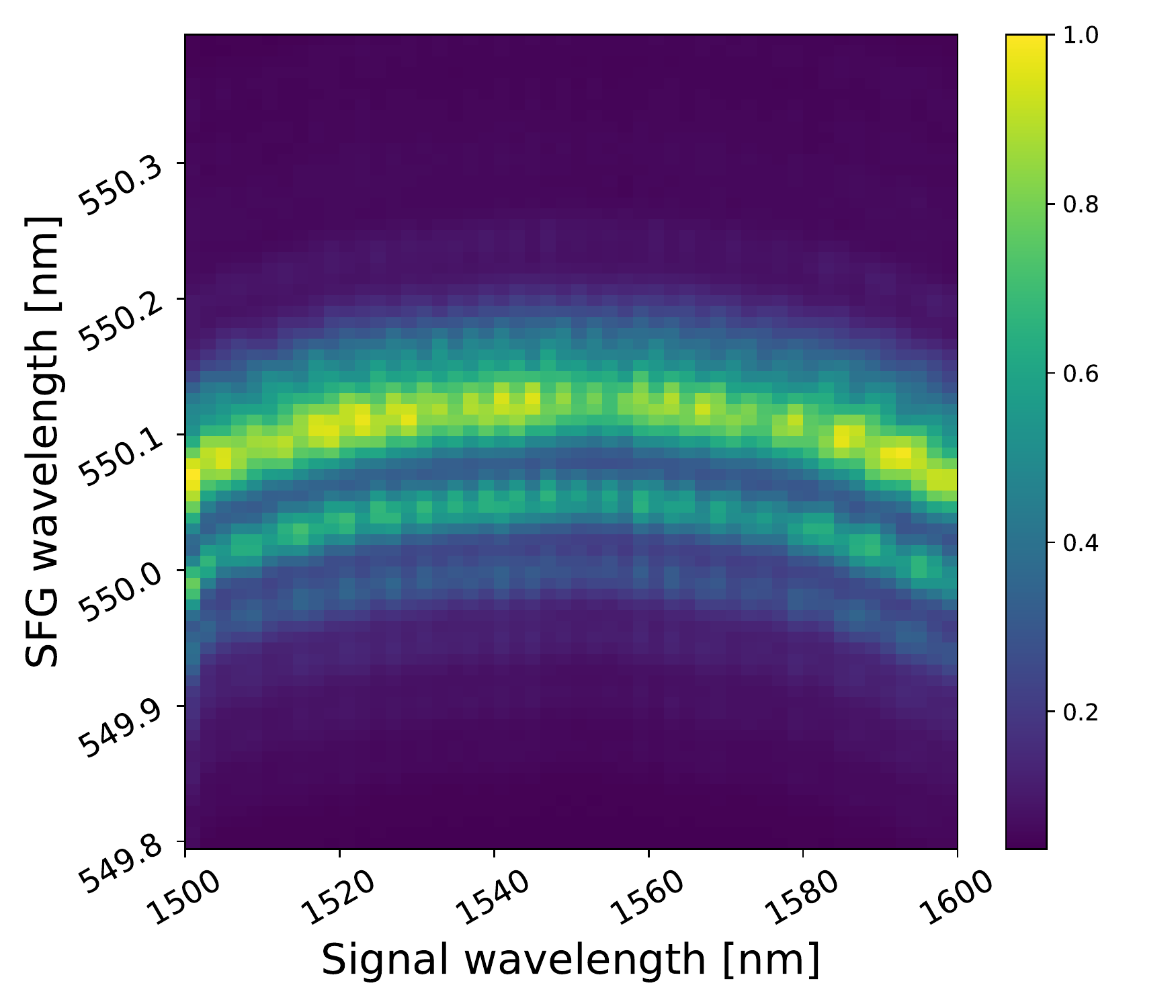}\quad
    \label{subimg:meas}}
	\subfloat[Simulated phasematching for a noisy waveguide.]{
    \includegraphics[width=.3\textwidth]{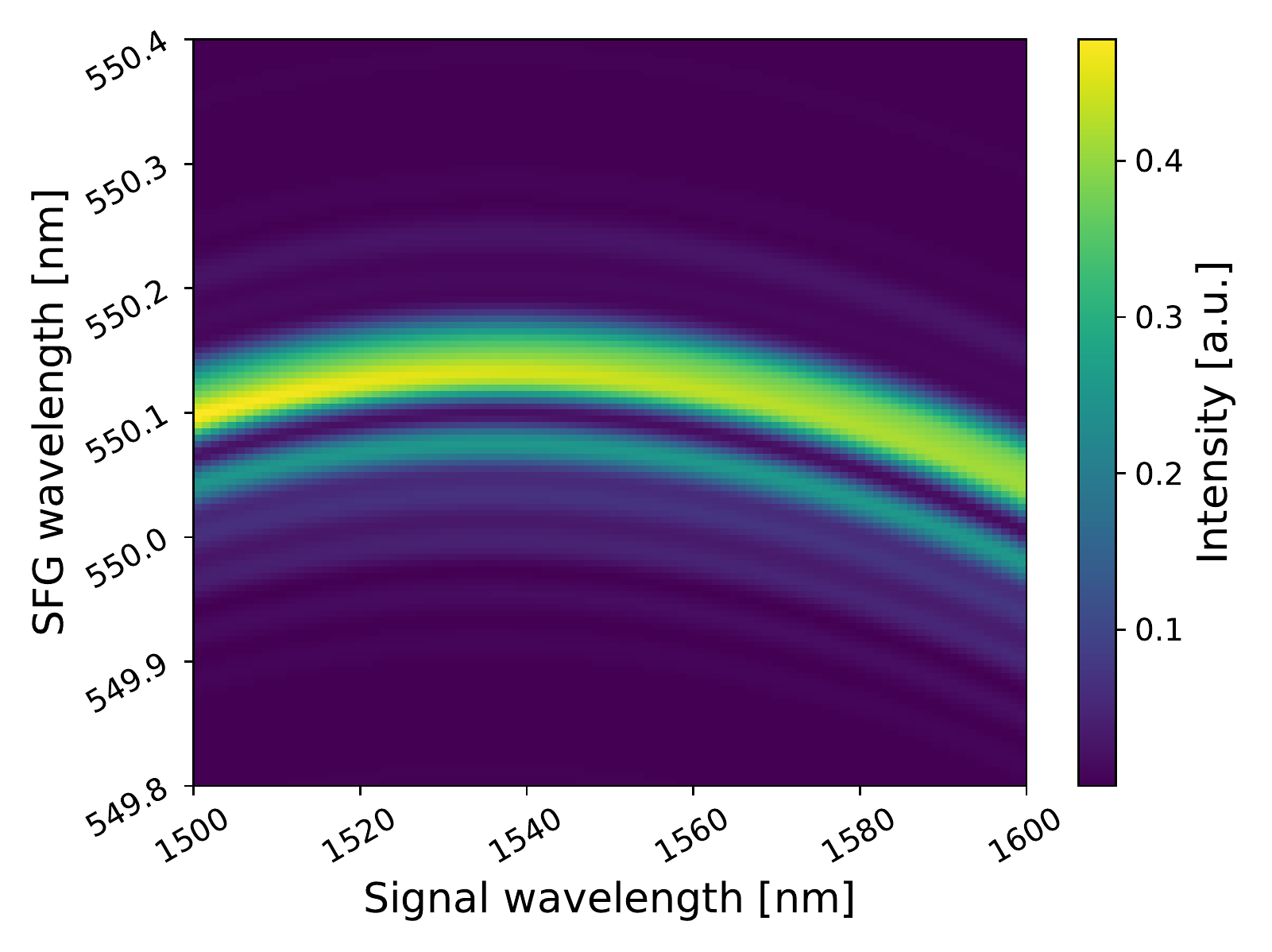}    
    \label{subimg:noisy_qpg}}
    \caption{Theoretical versus experimental performance of the bandwidth compression device presented in \cite{Allgaier2017}. In Figure \Ref{subimg:noiseless_qpg} the ideal phasematching is shown and can be compared with the measured phasematching shown in Figure \Ref{subimg:meas}. A phasematching similar to the measured one can be produced assuming a $1/f$ noise on the waveguide width and a maximum width error $\delta w$ = 0.4$\mu$m, as shown in Figure \Ref{subimg:noisy_qpg}.}
	\label{img:sim_JSA}
\end{figure}

Allgaier \textit{et al.} \cite{Allgaier2017} also characterized the phasematching spectrum of their device and the measurement showed deviations from the expected sinc$^{2}$ profile, as shown in Figure \ref{img:sim_JSA}
These deviations indicates the presence of non-negligible fabrication imperfections, as it was shown in Figure \ref{img:comparison_awgn_pink} that the presence of $1/f$ noise leads to more prominent side lobes and an asymmetric phasematching profile.
Assuming a $1/f$-noise profile and $\delta w = 0.4\mu$m, simulations have been able to reproduce the asymmetry and the prominent side lobes present in the measured phasematching, as can be seen comparing the measured and simulated spectra in Figure \ref{subimg:noisy_qpg} and Figure \ref{subimg:meas} respectively.

The limited compression and the shape of the phasematching lead us to conclude that the presence of fabrication imperfections limits the performance of the device presented in \cite{Allgaier2017}. 
A previously discussed method to overcome these limitations is to design the process to be noncritically phasematched.
Although Figure \ref{img:sensitivity_LN} shows that noncritical phasematching cannot be found for the QPG, it is possible to reduce the process sensitivity by increasing the waveguide width.
Figure \ref{subimg:bw_compr_13um} shows the calculated BCFs for a 13$\mu$m-wide waveguide as the length and the magnitude of the noise are varied.
The results show a greatly reduced sensitivity to noise, as large fabrication errors have a lower impact on the BCF.
These systems would reliably permit BCFs above 40, resulting in pulses with bandwidths below 25 GHz. 
Interestingly,  these bandwidths correspond to pulses longer than 20ps at 550nm, a regime that is often difficult to reach.
These results highlight the fact that proper system design can have a drastic impact on the final performance of such devices.

\section{Discussion}
\label{sec:discussion}
So far, we have shown that fabrication imperfections in integrated nonlinear systems limit the useful length and maximum efficiency of the devices.
A natural question is then how to overcome these limitations and optimize device performance.
Three general methods can be employed to improve the overall efficiency of the devices: reducing the magnitude of fabrication imperfections, designing the process to be insensitive to fabrication errors and reducing the impact of fabrication imperfections by using shorter waveguides.

Whilst it may be possible to reduce the magnitude of fabrication imperfections for a given production process, these devices are often realized using state-of-the-art technology and it may not be possible to make further improvements. 
In any case, the unavoidable presence of technological errors during waveguide production will impose an ultimate limit to device performance. 
It is therefore crucial to devise other methods to overcome these limits.

One solution is to design the process to reduce the sensitivity to fabrication errors, as discussed in \ref{subsec:mathdescription}. 
Ideally, one would design waveguides in an noncritically phasematched regime \cite{Lim1990}; however, this is not always possible, as shown in Figure \ref{img:sensitivity_LN}. 
Nevertheless, choosing the appropriate design will minimize the process sensitivity and reduce the impact of fabrication imperfections, as discussed at the end of section \ref{subsec:QPG}.

Another method to reduce the sensitivity to fabrication errors is to choose shorter waveguides; an unavoidable drawback is the reduction of the conversion efficiency.
This can be compensated for by employing multi-pass schemes, e.g. double-pass systems or cavity configurations. 
This technique, well suited for CW systems \cite{Stefszky2017}, removes the need for long waveguides at the cost of increased device complexity.

\section{Conclusions}
\label{sec:conclusions}
In this paper we derived a framework to study the limits posed by fabrication imperfections on the nonlinear performance of waveguide devices.
A qualitative model was first developed to describe the effect of fabrication imperfections on the performance of nonlinear waveguides.  
This model showed that long waveguides are more susceptible to fabrication errors occurring during the production.
A quantitative model was then introduced, which is able to account for inhomogeneities along the length of the waveguide. 
Applying this model to Ti:LN waveguides revealed that fabrication imperfections with long range spatial correlations leads to a reduced conversion efficiency and a distortion of the phasematching spectrum. 
Finally, we studied the impact of imperfections on three prominent quantum processes. 
Despite each process having different figures of merit, e.g. conversion efficiency or phasematching bandwidth, the performance of each process was found to be limited by the presence of fabrication imperfections.
Therefore, it is crucial to take fabrication imperfections into account when designing quantum optics devices.

\section*{Acknowledgement}
The authors thank J. M. Donohue and C. Eigner for valuable discussion and helpful comments.
This research has received funding from the European Union (EU) Horizon 2020 Research and Innovation program under Grant Agreement No. 665148.

\clearpage

\appendix
\section{Sensitivity to noise of the processes studied in Figure 3.}
\label{app:complete_processes}
\begin{figure}[bp]
    \centering
    \subfloat[Type-II decorrelated PDC]
    {
    \includegraphics[width=.4\textwidth]{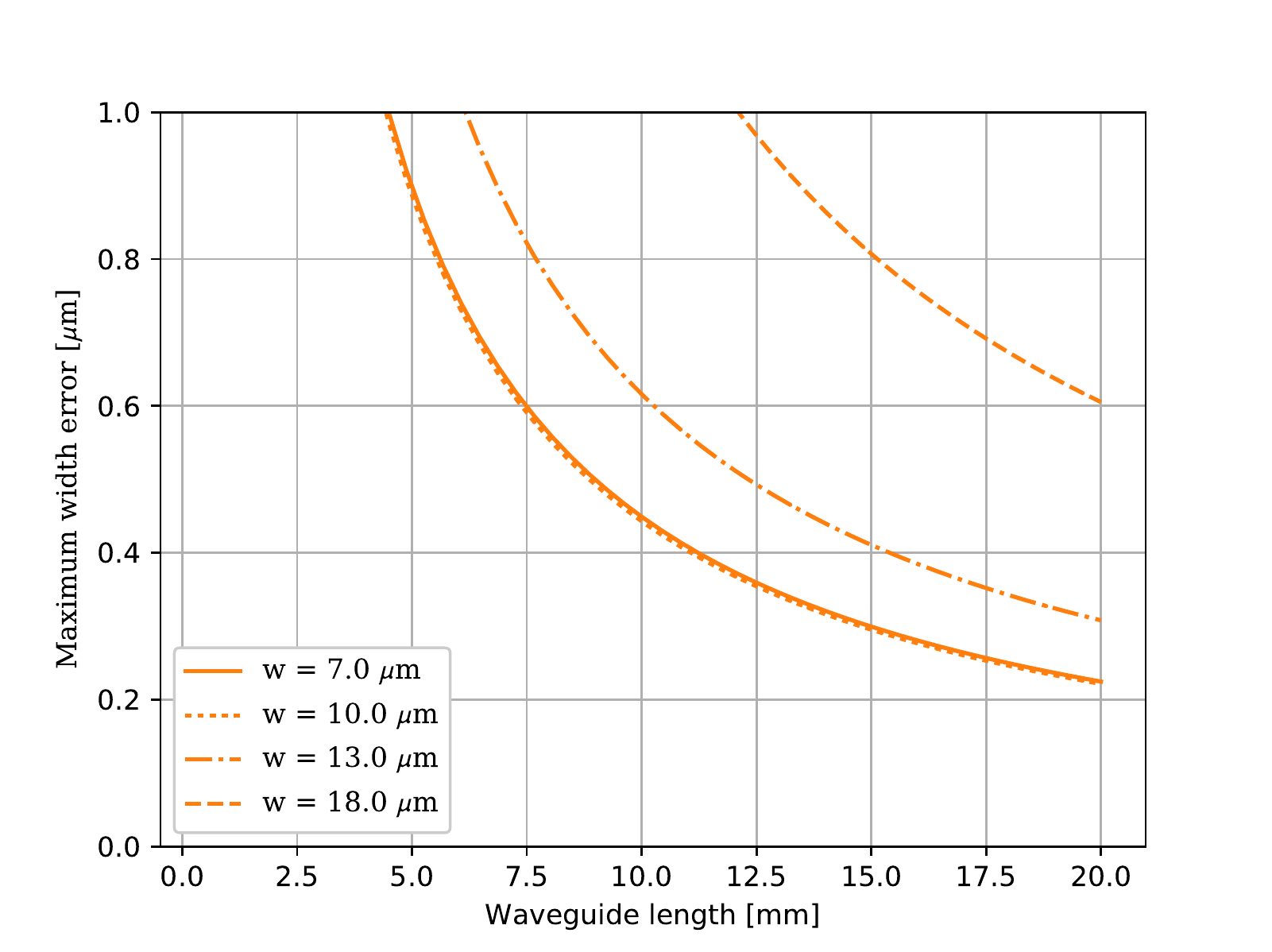}
    \label{subpl:LNSHGII}} \qquad
    \subfloat[Resonant PDC]
    {\includegraphics[width=.4\textwidth]{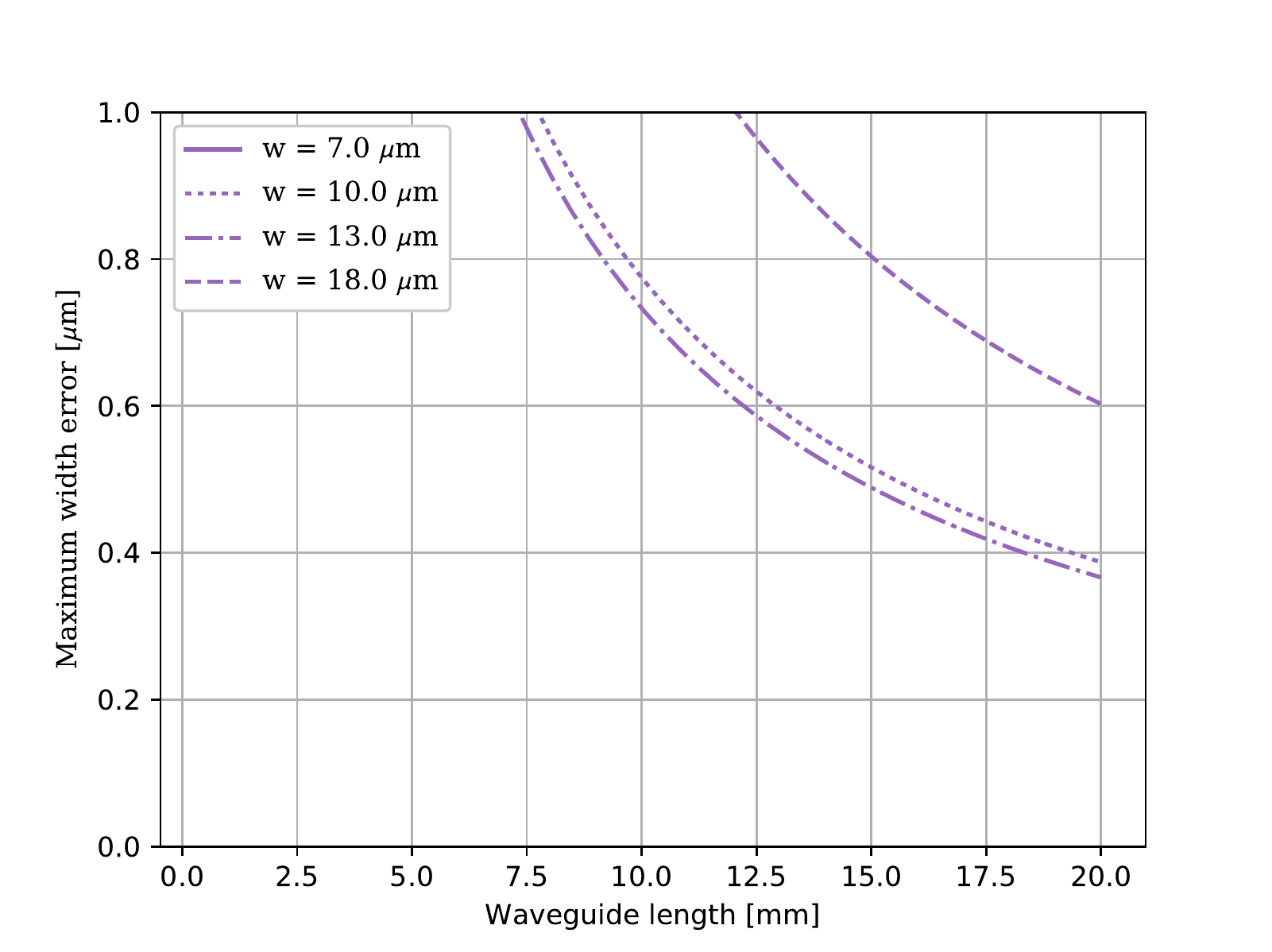}
    \label{subpl:LN_resonant}}\\
    \subfloat[Counter-propagating PDC]
    {\includegraphics[width=.4\textwidth]{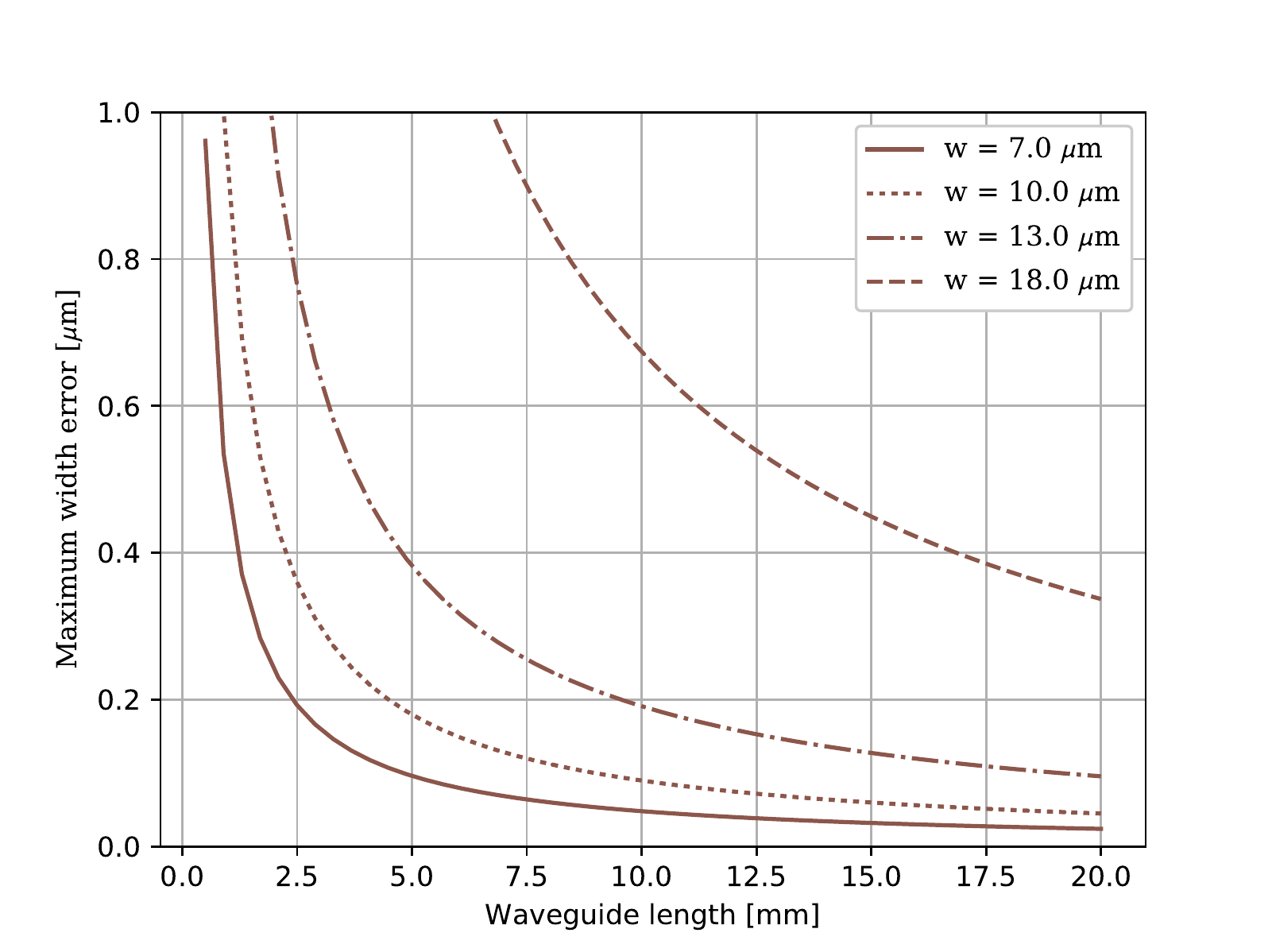}
    \label{subpl:LNBW1550}}
    \caption{Analysis of the sensitivity for different processes the process analysed in Figure 3 but not shown in Figure 4. The curve for $w=7\mu$m is missing from Figures (\ref{subpl:LN_resonant}) since the system is non-critically phasematched and therefore insensitive to noise.}
    \label{appendix_img:LNsensit}
\end{figure}

In Figure \ref{appendix_img:LNsensit} we report the relation between the maximum width error and the waveguide length for different waveguide widths, for the processes analysed in Figure 3 but not shown in Figure 4.
The three processes show very different behaviours. 
Counter-propagating PDC exhibits a very high sensitivity to waveguide width error, therefore the implementation of this process in LN waveguide will be particularly challenging.
On the other hand, the resonant PDC source shows a very low sensitivity to fabrication errors, suggesting that samples much longer than 2 cm could be produced reliably.
These simulations show the dramatic differences in process sensitivities of different devices, again highlighting the need for system design that accounts for the effect of fabrication errors.

\section{Mathematical models of the waveguide noise}
\setcounter{section}{1}
\label{app:mathmodelnoise}
To simulate the effect of errors on the waveguide width, knowledge of the effective refractive index $n_{eff}$ of the waveguide for variable waveguide widths and for different wavelengths is required. 
Therefore, using a FEM solver implemented in Python we computed the values of $n_{eff}$ for waveguide widths $w$ in the range $[5.5, 22]\mu$m in steps of 0.5$\mu$m, in the wavelength range from 400nm to 1700nm for both TE and TM polarizations. 
These values are used to calculate 
\begin{equation}
\Delta\beta(z) =  2\pi\left\lbrace\frac{ n_{eff}(\lambda_3, z)}{\lambda_3} -\frac{ n_{eff}(\lambda_2, z)}{\lambda_2} -\frac{n_{eff}(\lambda_1, z)}{\lambda_1} \right\rbrace
\end{equation}
as a function of the waveguide width for the different processes analysed.  

To simulate a single instance of a waveguide of length $L$, a mesh of $N$ points spaced by $\Delta z$ = 50 $\mu$m is first generated. 
A random waveguide profile with the desired noise spectrum is then generated as follows:
\begin{enumerate}
\item Create a vector of spatial frequencies $f_k = -\frac{1}{2\Delta z} \ldots \frac{1}{2\Delta z}$ in steps of $\Delta f=\frac{1}{L}$.
\item Generate the spectrum of the noise $C_k=\frac{1}{f_k^\gamma}\rme^{\rmi\phi_k}$ for $k\in\left[\frac{N}{2}, \frac{N}{2}\right]$, where $\gamma=0$ for AWG noise and $\gamma =1$ for pink noise and $\phi_k$ is a random variable uniformly distributed in $[0, 2\pi]$. In order to ensure a real-valued noise profile, we pose the condition $\phi_k = -\phi_{-k}$.
\item Take the inverse fast Fourier transform of the spectrum $\left\lbrace C_k\right\rbrace$ and normalise it to have a mean value of $w$ and maximum deviation of $|\delta w|$.
\end{enumerate}
Different random instances of the same noise spectrum are generated simply by randomly sampling the phase $\phi_k$.
Once the width profile $w(z)$ is generated, we calculate the spatially-dependent momentum mismatch $\Delta\beta(z)$ using the width-dependent Sellmeier equations previously derived.
Finally,  equation 
\begin{equation}
\phi \propto \frac{1}{L}\int_0^L \rme^{\rmi\int_0^z\Delta\beta(\xi) \rmd\xi} \rmd z.
\end{equation}
is discretised as
\begin{equation}
	\phi = \frac{\Delta z}{L}\sum_{n=0}^N\rme^{\rmi\Delta z\sum_{m=0}^n\left(\Delta\beta_m - \frac{2\pi}{\Lambda}\right)}
	\label{eq:discretization}
\end{equation}
and the phasematching is evaluated by computing $\Delta\beta(z_m)$ at each mesh point $z_m$, given the local waveguide width $w(z_m)$. 
Note that (\ref{eq:discretization}) already provides the phasematching spectrum normalised per unit length and $|\phi|^2$ will always have a maximum value of 1 (in the case of ideal phasematching) or lower.

\clearpage

\begin{thebibliography}{10}

\bibitem{Kumar1990}
Prem Kumar.
\newblock {Quantum frequency conversion}.
\newblock {\em Optics Letters}, 15(24):1476, 1990.

\bibitem{Chou1999}
M~H Chou, K~R Parameswaran, Martin~M Fejer, and I~Brener.
\newblock {Multiple-channel wavelength conversion by use of engineered
  quasi-phase-matching structures in LiNbO(3) waveguides.}
\newblock {\em Optics letters}, 24(16):1157--1159, 1999.

\bibitem{Cerullo2003}
Giulio Cerullo and Sandro {De Silvestri}.
\newblock {Ultrafast optical parametric amplifiers}.
\newblock {\em Review of Scientific Instruments}, 74(1 I):1--18, 2003.

\bibitem{Pysher2008}
Matthew Pysher, Russell Bloomer, Olivier Pfister, Christopher~M. Kaleva,
  Tony~D. Roberts, and Philip~R. Battle.
\newblock {Broadband amplitude squeezing in a periodically poled KTiOPO{\_}4
  waveguide}.
\newblock {\em Optics Letters}, 34(3):256--258, 2008.

\bibitem{Thyagarajan2009}
K.~Thyagarajan, J.~Lugani, S.~Ghosh, K.~Sinha, A.~Martin, D.~B. Ostrowsky,
  O.~Alibart, and S.~Tanzilli.
\newblock {Generation of polarization-entangled photons using type-II doubly
  periodically poled lithium niobate waveguides}.
\newblock {\em Physical Review A - Atomic, Molecular, and Optical Physics},
  80(5):1--8, 2009.

\bibitem{Stefszky2017}
Micheal Stefszky, Raimund Ricken, C.~Eigner, V.~Quiring, H.~Herrmann, and
  C.~Silberhorn.
\newblock {Waveguide Cavity Resonator as a Source of Optical Squeezing}.
\newblock {\em Physical Review Applied}, 7(4):1--5, 2017.

\bibitem{Albota2004}
Marius~A. Albota and Franco N.~C. Wong.
\newblock {Efficient single-photon counting at 1550 $\mu$m by means of
  frequency upconversion}.
\newblock {\em Optics Letters}, 29(13):1449, 2004.

\bibitem{Roussev2004}
Rostislav~V Roussev, Carsten Langrock, Jonathan~R Kurz, and Martin~M Fejer.
\newblock {Periodically poled lithium niobate waveguide sum-frequency generator
  for efficient single-photon detection at communication wavelengths}.
\newblock {\em Optics letters}, 29(13):1518--20, 2004.

\bibitem{Vandevender2004}
Aaron~P. Vandevender and Paul~G. Kwiat.
\newblock {High efficiency single photon detection via frequency
  up-conversion}.
\newblock {\em Journal of Modern Optics}, 51(9-10):1433--1445, 2004.

\bibitem{Pelc2012a}
J.~S. Pelc, Leo Yu, Kristiaan {De Greve}, Peter~L McMahon, Chandra~M Natarajan,
  Vahid Esfandyarpour, Sebastian Maier, Christian Schneider, Martin Kamp, Sven
  H{\"{o}}fling, Robert~H Hadfield, Alfred Forchel, Yoshihisa Yamamoto, and
  Martin~M Fejer.
\newblock {Downconversion quantum interface for a single quantum dot spin and
  1550-nm single-photon channel}.
\newblock {\em Optics Express}, 20(25):27510, 2012.

\bibitem{Rutz2017}
Helge R{\"{u}}tz, Kai~Hong Luo, Hubertus Suche, and Christine Silberhorn.
\newblock {Quantum Frequency Conversion between Infrared and Ultraviolet}.
\newblock {\em Physical Review Applied}, 7(2):1--7, 2017.

\bibitem{Maring2017}
Nicolas Maring, Pau Farrera, Kutlu Kutluer, Margherita Mazzera, Georg Heinze,
  and Hugues de~Riedmatten.
\newblock {Photonic quantum state transfer between a cold atomic gas and a
  crystal}.
\newblock {\em Nature}, 551(7681):485--488, 2017.

\bibitem{Orieux2016}
Adeline Orieux and Eleni Diamanti.
\newblock {Recent advances on integrated quantum communications}.
\newblock {\em Journal of Optics}, 18(8):083002, 2016.

\bibitem{Montaut2017}
Nicola Montaut, Linda Sansoni, Evan Meyer-Scott, Raimund Ricken, Viktor
  Quiring, Harald Herrmann, and Christine Silberhorn.
\newblock {High-Efficiency Plug-and-Play Source of Heralded Single Photons}.
\newblock {\em Phys. Rev. Applied}, 8(2):24021, aug 2017.

\bibitem{Krapick2013}
Stephan Krapick, Harald Herrmann, Benjamin Brecht, Viktor Quiring, Hubertus
  Suche, and Christine Silberhorn.
\newblock {A highly efficient integrated two-color source for heralded single
  photons}.
\newblock {\em 2013 Conference on Lasers and Electro-Optics Europe and
  International Quantum Electronics Conference, CLEO/Europe-IQEC 2013}, 2013.

\bibitem{Kruse2015}
Regina Kruse, Linda Sansoni, Sebastian Brauner, Raimund Ricken, Craig~S.
  Hamilton, Igor Jex, and Christine Silberhorn.
\newblock {Dual-path source engineering in integrated quantum optics}.
\newblock {\em Physical Review A - Atomic, Molecular, and Optical Physics},
  92(5):1--6, 2015.

\bibitem{Sansoni2017}
Linda Sansoni, Kai~Hong Luo, Christof Eigner, Raimund Ricken, Viktor Quiring,
  Harald Herrmann, and Christine Silberhorn.
\newblock {A two-channel, spectrally degenerate polarization entangled source
  on chip}.
\newblock {\em npj Quantum Information}, 3(1):1--5, 2017.

\bibitem{Lenzini2018arxiv}
Francesco Lenzini, Jiri Janousek, Oliver Thearle, Matteo Villa, Ben Haylock,
  Sachin Kasture, Liang Cui, Hoang-Phuong Phan, Dzung~Viet Dao, Hidehiro
  Yonezawa, Koy Lam, Elanor~H Huntington, and Mirko Lobino.
\newblock {Integrated photonic platform for quantum information with continuous
  variables}.

\bibitem{Lim1990}
E.~J. Lim, S.~Matsumoto, and Martin~M Fejer.
\newblock {Noncritical phase matching for guided-wave frequency conversion}.
\newblock {\em Applied Physics Letters}, 57(22):2294--2296, 1990.

\bibitem{Bortz1994}
M~L Bortz, S~J Field, Martin~M Fejer, D~W Nam, R~G Waarts, and D~F Welch.
\newblock {Noncritical Quasi-Phase-Matched Second Harmonic Generation in an
  Annealed Proton-Exchanged LiNbO3 Waveguide}.
\newblock {\em IEEE Transactions on Quantum Electronics}, 30(12):2953--2960,
  1994.

\bibitem{Pelc2010}
J.~S. Pelc, C~Langrock, Qiang Zhang, and Martin~M Fejer.
\newblock {Influence of domain disorder on parametric noise in
  quasi-phase-matched quantum frequency converters.}
\newblock {\em Optics letters}, 35(16):2804--2806, 2010.

\bibitem{Pelc2011a}
J.~S. Pelc, C~R Phillips, D~Chang, C~Langrock, and Martin~M Fejer.
\newblock {Efficiency pedestal in quasi-phase-matching devices with random
  duty-cycle errors.}
\newblock {\em Optics letters}, 36(6):864--866, 2011.

\bibitem{Phillips2013}
C~R Phillips, J.~S. Pelc, and Martin~M Fejer.
\newblock {Parametric processes in quasi-phasematching gratings with random
  duty cycle errors}.
\newblock {\em Journal of the Optical Society of America B}, 30(4):982--993,
  2013.

\bibitem{FrancisJones2016}
Robert J.~A. Francis-Jones and Peter~J. Mosley.
\newblock {Characterisation of longitudinal variation in photonic crystal
  fibre}.
\newblock {\em Optics Express}, 24(22):24836, 2016.

\bibitem{Boyd2008}
Robert~W. Boyd.
\newblock {\em {Nonlinear Optics}}.
\newblock Academic Press, Inc., Orlando, FL, USA, 3rd edition, 2008.

\bibitem{Suhara2003}
Toshiaki Suhara and Masatoshi Fujimura.
\newblock {\em {Waveguide Nonlinear-Optic Devices}}.
\newblock Springer-Verlag Berlin Heidelberg, 1 edition, 2003.

\bibitem{Regener1988}
R.~Regener and Wolfgang Sohler.
\newblock {Efficient second-harmonic generation in Ti:LiNbO{\_}3 channel
  waveguide resonators}.
\newblock {\em Journal of the Optical Society of America B}, 5(2):267, 1988.

\bibitem{Amin1997}
J~Amin, V~Pruneri, J~Webj{\"{o}}rn, P.~St.~J Russell, D.~C Hanna, and J.~S
  Wilkinson.
\newblock {Blue light generation in a periodically poled Ti:LiNbO3 channel
  waveguide}.
\newblock {\em Optics Communications}, 135(1–3):41--44, 1997.

\bibitem{Kanbara1999}
Hirohisa Kanbara, Hiroki Itoh, Masaki Asobe, and Kazuto Noguchi.
\newblock {All-Optical Switching Based on Cascading of Second-Order
  Nonlinearities in a Periodically Poled Titanium-Diffused Lithium Niobate
  Waveguide}.
\newblock {\em IEEE Photonics Technology Letters}, 11(3):328--330, 1999.

\bibitem{Eckstein2011}
Andreas Eckstein, Benjamin Brecht, and Christine Silberhorn.
\newblock {A quantum pulse gate based on spectrally engineered sum frequency
  generation}.
\newblock {\em Optics Express}, 19(15):13770, 2011.

\bibitem{Luo2015}
Kai~Hong Luo, Harald Herrmann, Stephan Krapick, Benjamin Brecht, Raimund
  Ricken, Viktor Quiring, Hubertus Suche, Wolfgang Sohler, and Christine
  Silberhorn.
\newblock {Direct generation of genuine single-longitudinal-mode narrowband
  photon pairs}.
\newblock {\em New Journal of Physics}, 17(7), 2015.

\bibitem{Stefszky2018}
Micheal Stefszky, R.~Ricken, C.~Eigner, V.~Quiring, H.~Herrmann, and
  C.~Silberhorn.
\newblock {High-power waveguide resonator second harmonic device with external
  conversion efficiency up to 75{\%}}.
\newblock {\em Journal of Optics (United Kingdom)}, 20(6), 2018.

\bibitem{Strake1988}
E.~Strake, G.~P. Bava, and I.~Montrosset.
\newblock {Guided Modes of Ti:LiNbO3 Channel Waveguides: A Novel
  Quasi-Analytical Technique in Comparison with the Scalar Finite-Element
  Method}.
\newblock {\em Journal of Lightwave Technology}, 6(6):1126--1135, 1988.

\bibitem{Allgaier2017}
Markus Allgaier, Vahid Ansari, Linda Sansoni, Christof Eigner, Viktor Quiring,
  Raimund Ricken, Georg Harder, Benjamin Brecht, and Christine Silberhorn.
\newblock {Highly efficient frequency conversion with bandwidth compression of
  quantum light}.
\newblock {\em Nature Communications}, 8:1--6, 2017.

\bibitem{Helmfrid1992}
Sten Helmfrid, Gunnar Arvidsson, and Jonas Webj{\"{o}}rn.
\newblock {Influence of various imperfections on the conversion efficiency of
  second-harmonic generation in quasi-phase-matching lithium niobate
  waveguides}.
\newblock {\em Journal of Opt. Soc. Am. B}, 10(2):222--229, 1992.

\bibitem{Helmfrid1991}
Sten Helmfrid and Gunnar Arvidsson.
\newblock {Influence of randomly varying domain lengths and nonuniform
  effective index on second-harmonic generation in quasi-phase-matching
  waveguides}.
\newblock {\em Journal of Opt. Soc. Am. B}, 8(4), 1991.

\bibitem{Farahmand2004}
Mitra Farahmand and Martijn de~Sterke.
\newblock {Parametric amplification in presence of dispersion fluctuations.}
\newblock {\em Optics Express}, 12(1):136--142, 2004.

\bibitem{Menicucci2006}
Nicolas~C. Menicucci, Peter {Van Loock}, Mile Gu, Christian Weedbrook,
  Timothy~C. Ralph, and Michael~A. Nielsen.
\newblock {Universal quantum computation with continuous-variable cluster
  states}.
\newblock {\em Physical Review Letters}, 97(11):13--16, 2006.

\bibitem{Bowen2002}
Warwick~P. Bowen, Nicolas Treps, Roman Schnabel, and Ping~Koy Lam.
\newblock {Experimental Demonstration of Continuous Variable Polarization
  Entanglement}.
\newblock {\em Physical Review Letters}, 89(25):1--4, 2002.

\bibitem{Yukawa2008}
Mitsuyoshi Yukawa, Ryuji Ukai, Peter~Van Loock, and Akira Furusawa.
\newblock {Experimental generation of four-mode continuous-variable cluster
  states}.
\newblock {\em Physical Review A}, 78(1):012301, 2008.

\bibitem{Aasi2013}
J.~Aasi, J.~Abadie, B.~P. Abbott, et al.
\newblock {Enhanced sensitivity of the LIGO gravitational wave detector by
  using squeezed states of light}.
\newblock {\em Nature Photonics}, 7(8):613--619, 2013.

\bibitem{Serkland1995}
D.~K. Serkland, Martin~M Fejer, R~L Byer, and Y~Yamamoto.
\newblock {Squeezing in a quasi-phase-matched LiNbO{\_}3 waveguide.}
\newblock {\em Optics letters}, 20(15):1649--51, 1995.

\bibitem{Olislager2010}
L.~Olislager, J.~Cussey, A.~T. Nguyen, P.~Emplit, S.~Massar, J.~M. Merolla, and
  K.~Phan Huy.
\newblock {Frequency-bin entangled photons}.
\newblock {\em Physical Review A - Atomic, Molecular, and Optical Physics},
  82(1):1--7, 2010.

\bibitem{Zhong2015}
Tian Zhong, Hongchao Zhou, Robert~D. Horansky, Catherine Lee, Varun~B. Verma,
  Adriana~E. Lita, Alessandro Restelli, Joshua~C. Bienfang, Richard~P. Mirin,
  Thomas Gerrits, Sae~Woo Nam, Francesco Marsili, Matthew~D. Shaw, Zheshen
  Zhang, Ligong Wang, Dirk Englund, Gregory~W. Wornell, Jeffrey~H. Shapiro, and
  Franco N.~C. Wong.
\newblock {Photon-efficient quantum key distribution using time-energy
  entanglement with high-dimensional encoding}.
\newblock {\em New Journal of Physics}, 17, 2015.

\end{thebibliography}

\end{document}